\documentclass[twocolumn,showpacs,superscriptaddress,preprintnumbers,amsmath,amssymb,aps,prx]{revtex4-1}
\usepackage{graphicx}
\usepackage{mathptmx}
\usepackage{verbatim}
\usepackage{graphicx}
\usepackage{amsmath}
\usepackage{amssymb}
\usepackage{amsfonts}
\usepackage{mathrsfs}
\usepackage{amsmath}     
\usepackage{color}
\usepackage{bm}               
\usepackage{amssymb}
\usepackage{xspace} 
\usepackage{hyperref}
\usepackage{multirow}
\usepackage{color}
\usepackage{float}
\usepackage{bbold}
\usepackage{tabu}
\usepackage{textcomp}
\usepackage{multirow}
\usepackage{dcolumn}
\usepackage{appendix} 

\newcolumntype{L}[1]{>{\raggedright\arraybackslash}m{#1}}
\newcolumntype{C}[1]{>{\centering\arraybackslash}m{#1}}
\newcolumntype{R}[1]{>{\raggedleft\arraybackslash}m{#1}}
\newcolumntype{N}{@{}m{0pt}@{}}

\usepackage{listings}
\usepackage{color}
\usepackage[utf8]{inputenc}

\definecolor{dkgreen}{rgb}{0,0.6,0}
\definecolor{gray}{rgb}{0.5,0.5,0.5}
\definecolor{mauve}{rgb}{0.58,0,0.82}

\begin{document}

\title{Relaxation Effects in Twisted Bilayer Graphene:\\a Multi-Scale Approach}

\author{Nicolas Leconte}
\affiliation{Department of Physics, University of Seoul, Seoul 02504, Korea}
\author{Srivani Javvaji}
\affiliation{Department of Physics, University of Seoul, Seoul 02504, Korea}
\author{Jiaqi An}
\affiliation{Department of Physics, University of Seoul, Seoul 02504, Korea}
\affiliation{Department of Smart Cities, University of Seoul, Seoul 02504, Korea}
\author{Appalakondaiah Samudrala}
\affiliation{Department of Physics, University of Seoul, Seoul 02504, Korea}
\author{Jeil Jung}
\email{jeiljung@uos.ac.kr}
\affiliation{Department of Physics, University of Seoul, Seoul 02504, Korea}
\affiliation{Department of Smart Cities, University of Seoul, Seoul 02504, Korea}

\date{\today}
\begin{abstract}
We present a multi-scale density functional theory (DFT) informed molecular dynamics 
and tight-binding (TB) approach to capture the interdependent atomic and electronic 
structures of twisted bilayer graphene. 
We calibrate the flat band magic angle to be at $\theta_{\rm M} = 1.08^{\circ}$ 
by rescaling the interlayer tunneling for different atomic structure relaxation models 
as a way to resolve the indeterminacy of existing atomic and electronic structure models whose 
predicted magic angles vary widely between $0.9^\circ \sim 1.3^\circ$.
The interatomic force fields are built using input from various stacking and interlayer distance dependent
DFT total energies including the exact exchange and random phase approximation (EXX+RPA). 
We use a Fermi velocity of $\upsilon_{\rm F} \simeq 10^{6}$~m/s for graphene
that is enhanced by about $\sim 15\%$ over the local density approximation (LDA) values.
Based on this atomic and electronic structure model we obtain high-resolution spectral functions 
comparable with experimental angle-resolved photoemission spectra (ARPES).
Our analysis of the interdependence between the atomic and electronic structures
indicates that the intralayer elastic parameters compatible with the DFT-LDA, 
which are stiffer by $\sim$30\% than widely used reactive empirical bond order force fields, 
can combine with EXX+RPA interlayer potentials to yield the magic angle at $\sim 1.08^{\circ}$ 
without further rescaling of the interlayer tunneling.
\end{abstract}
\maketitle

\section{Introduction}

%
The discovery of correlated insulating phases and superconductivity~\cite{cao2018} 
in twisted bilayer graphene (tBG) has boosted the field of twistronics~\cite{Bistritzer:2011ho, LopesdosSantos:2007fg} 
where strong electron-electron interactions~\cite{Yankowitz2019, Kerelsky2019,choi2019} 
play a dominant role in the physics at specific magic angles where the bands become nearly flat. 
Experimental magic angle values are reported within varying ranges due to their sensitivity to the cleanliness 
of the sample affecting the Fermi velocity and the strength of electron-electron interaction effects~\cite{Kerelsky2019,choi2019}.
Existing electronic structure models~\cite{laissardire2012, Morell2010, nam2017, Uchida2014} have been refined~\cite{Angeli2018, Guinea2019} 
to understand the peculiarities of the physics at play at these specific magic angles. 
Theoretically, the magic angles depend on the chosen model Hamiltonian. 
Of particular importance are the relaxation effects in van der Waals heterostructures~\cite{woods, Jung2015} 
that have already been reported using a variety of methods including (i) fully atomic lattice relaxation approaches~\cite{Wijk2015, jain2016,nam2017,Lucignano2019}, 
(ii) non-linear finite element plate models~\cite{zhang2018}, (iii) a generalized-stacking fault energy (GSFE) analysis~\cite{Dai2016}, possibly combined with (iv) a configuration-space representation~\cite{carr2018}, and finally, very commonly, with (v) computationally non-prohibitive continuum models~\cite{carr2019a,Guinea2019, Walet_2019}. The common denominator in these analyses is the observation of (i) a reduction in the size of the AA stacking region, an increase of the AB/BA regions and the appearance of sharper stacking domain walls with decreasing twist angle and (ii) the tendency to lock the rotational alignment between the layers at the AA stacked regions for small twist angles. 
These mechanical effects contribute to the formation of secondary isolation gaps of the flat bands 
from higher energy bands~\cite{nam2017,carr2018, PhysRevX.8.031087}, 
enhance the value of the first magic angle and broadens the bands of the magic angles below $1^{\circ}$~\cite{carr2019a}. 
Quantitative conclusions inferred from electronic band structure (EBS) and spectral function plots depend on the aproximations used. 
%

Here, we propose an approach to 
capture the interdependent atomic and electronic structures of twisted bilayer graphene (tBG) 
by calibrating the predicted magic angle to  
the experimental value of $\theta_{\rm M} = 1.08^{\circ}$ and resolve the indeterminacy of 
the models in the literature for different atomic/electronic structure model 
combinations~\cite{laissardire2012, Fang:2016iq, Kolmogorov_2005, Gargiulo_2017, van_Wijk_2014, PhysRevB.98.235404, PhysRevB.87.041108} 
whose predicted magic angles vary widely between $0.9^\circ \sim 1.3^\circ$.
%
For this purpose we attempt a progressive refinement in the accuracy of our models.
For the relaxed atomic structure we use interatomic force fields based on dihedral registry-dependent interlayer 
potentials (DRIP)~\cite{wen2018} 
using parameter sets that reproduce the stacking registry and interlayer distance dependent total energies 
obtained within density functional theory.
Among the proposed parametrizations we have the systematically improved exact exchange and random 
phase approximation (EXX+RPA)~\cite{PhysRevB.96.195431}
which predicts structural reconstructions that are slightly weaker for the out-of-plane 
corrugation amplitudes than in commonly reported force field 
calculations~\cite{carr2018,nam2017,carr2019a, Gargiulo_2017, wen2018, Kolmogorov_2005}.
Comparison of in-plane relaxations against LDA-parametrized force fields give results that are similar 
to EXX-RPA-parametrized results due to similar energy differences~\cite{PhysRevB.96.195431} 
thus yielding a similarly strong driving force for unfavorable AA to favorable AB stacking rearrangement. 
%

The electronic structure model relies on rescaled interlayer hopping terms
for each atomic relaxation scheme to calibrate the flat band magic angle to be at $\sim 1.08^{\circ}$.
The interlayer hopping terms are modeled through widely used isotropic two-center (TC) distance 
dependent functions. We then propose an improvement of this model by replacing the intralayer terms using a strain-dependent version of the F2G2 model~\cite{jung2013}, by rescaling of the TC paremeters through interlayer tunneling values fitting at the Dirac point for all possible
stacking configurations obtained within LDA density functional theory (DFT) calculations~\cite{jung2014}, and by a relaxation-scheme dependent rescaling of the coupling strength to match the magic angle at a set value, coined as the Scaled Hybrid Exponential (SHE) model.
The specific magic angle value not only depends on the atomic structure but also on the 
intralayer Fermi velocity of graphene for which we use a value  $\upsilon_{\rm F} \simeq10^{6}$~m/s
that is enhanced by $\sim$15\% with respect to the local density approximation estimate 
$\upsilon_{\rm F} \simeq 0.84 \times10^{6}$~m/s.
%
%
%
%
%
We find that the interlayer tunneling does not require rescaling when 
local density approximation (LDA) elastic properties are used in combination with exact 
exchange random phase approximation (EXX-RPA) interlayer potentials. 
We further compare the freestanding tBG against the hexagonal boron nitride (hBN) supported 
electronic structure to confirm that the EBS, and therefore the magic angles values, are 
relatively insensitive to the substrate when the hBN has a large twist with respect to the contacting graphene layer. 

The manuscript is structured as follows.
In Sect.~\ref{methodology} we present details of the atomic structure calculations
through molecular dynamics simulations. In Sect.~\ref{TBSect} we discuss the details of the electronic bandstructure calculations using our SHE TB model while in Sect.~\ref{EBSSection} we provide results using this model.
In Sect.~\ref{spectralMethodSect}, we focus on the spectral function methods to illustrate possible signatures for three relevant tBG systems.
In Section~\ref{conclusion} we summarize our main findings.

\begin{figure*}[th]
    \includegraphics[width=0.9\textwidth]{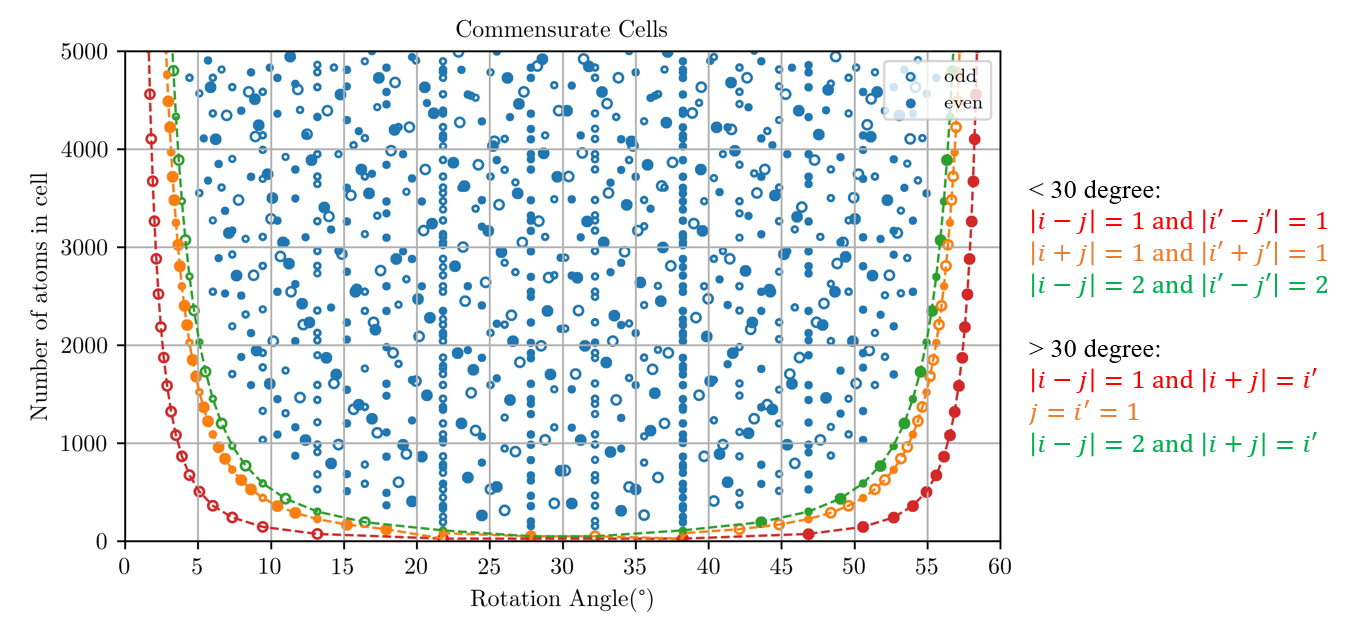}
    \caption{(color online) Number of atoms $N$ in the supercell as a function of twist angle obtained by 
    allowing different combinations of integers $i = j^{\prime}$, $j = i^{\prime}$ in Eq.~(\ref{twistangle}) and satisfying $\alpha=1$ in Eq.~(\ref{alphaEq}).
    We note that in the small angle approximation 
    the minimum of atoms in the supercell increases for small twist angles following the relation $N \simeq (i - j)^2/ \theta^2$.
    The hollow circle and filled circle point out the different parity for each dots.}
    \label{commensuration}
\end{figure*}

\section{Atomic structure calculations}
\label{methodology}
We begin by introducing our approach to obtain the relaxed atomic structure of twisted van der Waals systems. 
The structural relaxations rely on LAMMPS molecular dynamics (MD) simulation 
package~\cite{LAMMPS, Plimpton_1995} for which we use pairwise interlayer 
interaction force fields modeled to reproduce interlayer stacking dependent DFT total energies.
In the following we outline the method to obtain the commensurate supercell for different twist angles,
then we explain how to parametrize the pairwise potentials that reproduce different stacking 
dependent interlayer interaction energies including EXX+RPA,
and finally discuss the important role the choice in intralayer potential plays in correctly reproducing the elastic properties that govern the strength of the lattice reconstruction effects.
%
%
%
%
%
%

%
%

\subsection{Commensuration angles}
Commensurate supercells of twisted bilayer graphene can be formed  for a discrete set of twist angles $\theta$.
We use the procedure in Ref.~\cite{Hermann:2012dy} to relate a given twist angle between two 
rotated layers on top of each other using four integers
$i$, $j$, $i^\prime$ and $j^\prime$ through the relation
\begin{equation}
\cos(\theta) = \frac{1}{2 \alpha g}\left[ 2 i^\prime i + 2 j^\prime j + i^\prime j + j^\prime i \right]
\label{twistangle}
\end{equation}
where the scaling factor $\alpha$ is the ratio between the 
lattice constants $a$ and $a^{\prime}$ of the bottom and top layers respectively
\begin{equation}
\alpha = \frac{a^{\prime}}{a}
   = \sqrt{\frac{(i^{\prime 2} +j^{\prime 2}  + i^\prime j^\prime )}{(i^{ 2} +j^{ 2}  + i j )}}
   \label{alphaEq}
\end{equation}
and
\begin{equation}
g = i^2 + j^2 + i j.
\end{equation}
The two lattice vectors of the commensurate supercell ${\bm r}_{1}$, ${\bm r}_{2}$ 
can be related with the lattice vectors of the bottom reference layer ${\bm a}_{1}$, ${\bm a}_{2}$
and the top twisted layer ${\bm a}^\prime_{1}$, ${\bm a}^\prime_{2}$ through
\begin{equation}
\begin{pmatrix}
{\bm r}_{1} \\
{\bm r}_{2}
\end{pmatrix}
= {\bm M}^{\prime} \cdot
\begin{pmatrix}
{\bm a}_{1} \\
{\bm a}_{2}
\end{pmatrix}
= {\bm M} \cdot
\begin{pmatrix}
{\bm a}^{\prime}_{1} \\
{\bm a}^{\prime}_{2} 
\end{pmatrix}
\end{equation}
where we use the transformation matrices
\begin{eqnarray}
{\bm M} &=&
\begin{pmatrix}
      i      &     j        \\
    - j      &     i + j
\end{pmatrix}, 
\quad
{\bm M}^{\prime} =
\begin{pmatrix}
     i^\prime      & j^\prime       \\
    -j^\prime     & i^\prime + j^\prime
\end{pmatrix}.
\end{eqnarray}

In tBG we have $\alpha = 1$ scaling factor because the lattice constants of both top 
and bottom layers are equal. 
For illustration purposes we comment on the subset of commensurate superlattices 
that are obtained by imposing $i = j^{\prime}$ and $j = i^{\prime}$ 
or alternatively $i = -j^{\prime}$ and $j = -i^{\prime}$ that automatically satisfies the equal
lattice constant condition. 
Note that the use of identical indices $i = i^{\prime}$ and $j = j^{\prime}$ leaves the layers unrotated without introducing any change and corresponds to a trivial case, 
and switching signs $i = -i^{\prime}$ and $j = -j^{\prime}$ rotates the layers by 60$^{\circ}$.
\begin{table}[th]
\begin{tabular}{|c|c|c|c|l|c|c|}
\hline
$\theta^{\circ}$  & $i = j^{\prime}$, $j = i^{\prime}$  & \# Atoms  &  &   
$\theta^{\circ}$  & $i = j^{\prime}$, $j = i^{\prime}$  & \# Atoms  \\ \hline
0.100 & ~ 331,~~  330 & 1310764 &  & 1.696 & ~ 20,~~  19 & 4564   \\ \hline
0.200 & ~ 166,~~  165 & 328684    &  & 2.005 & ~ 17,~~  16 & 3268 \\ \hline
0.300 & ~ 111,~~  110 & 146524    &  & 2.134 & ~ 16,~~  15 & 2884 \\ \hline 
0.400 & ~ 83,~~  82 & 81676       &  & 2.281 & ~ 15,~~  14 & 2524 \\ \hline
0.497 & ~ 67,~~  66 & 53068       &  & 2.450 & ~ 14,~~  13 & 2188 \\ \hline
0.596 & ~ 56,~~  55 & 36964       &  & 2.646 & ~ 13,~~  12 & 1876 \\ \hline
0.797 & ~ 42,~~  41 & 20668       &  & 2.876 & ~ 12,~~  11 & 1588 \\ \hline
0.987 & ~ 34,~~  33 & 13468       &  & 3.150 & ~ 11,~~  10 & 1324 \\ \hline
1.018 & ~ 33,~~  32 & 12676       &  & 3.481 & ~ 10,~~  9 & 1084  \\ \hline
1.050 & ~ 32,~~  31 & 11908       &  & 3.890 & ~ 9,~~  8 & 868  \\ \hline
1.085 & ~ 31,~~  30 & 11164       &  & 4.408 & ~ 8,~~  7 & 676  \\ \hline
1.121 & ~ 30,~~  29 & 10444       &  & 5.086 & ~ 7,~~  6 & 508  \\ \hline
1.539 & ~ 22,~~  21 & 5548        &  & 6.009 & ~ 6,~~  5 & 364  \\ \hline
1.614 & ~ 21,~~  20 & 5044        &  & 7.341 & ~ 5,~~  4 & 244  \\ \hline
1.696 & ~ 20,~~  19 & 4564        &  & 9.430 & ~ 4,~~  3 & 148  \\ \hline
1.788 & ~ 19,~~  18 & 4108        &  & 13.174 & ~ 3,~~ 2 & 76   \\ \hline
1.890 & ~ 18,~~  17 & 3676        &  & 21.787 & ~ 2,~~ 1 & 28   \\ \hline
\end{tabular}
\caption{Selection of twist angles, indices and number of atoms in the smallest commensurate 
supercells corresponding to the red dots in Fig.~\ref{commensuration}.}
\end{table}

The areas of the commensurate supercell $A$ and the moire cell $A_M$ are 
related by an integer multiple $m$ through
\begin{eqnarray}
 m  =  A / A_{\rm M} = {\rm det}({\bm M} - {\bm M}^{\prime}) 
\end{eqnarray}
where $m = (i-j)^2$ takes a simple form when $i = j'$ and $j = i'$.
The smallest supercells corresponding to area multiples of $m = 1, 2, 3$ between the supercell
and the moire cell are represented as red, orange, and green in Fig.~\ref{commensuration}. 
These three area ratios can be obtained using the nontrivial $i \neq i^{\prime}$ condition when
\begin{eqnarray}
\left|i-j\right| &=& 1 \enspace \& \enspace \left|i^\prime-j^\prime \right| = 1, 
\enspace {\rm for} \enspace m = 1   \label{minsize} \\
\left|i+j\right| &=& 1 \enspace \& \enspace \left|i^\prime+j^\prime \right| = 1, 
\enspace  {\rm for} \enspace m = 2 \\
\left|i-j\right| &=& 2 \enspace \& \enspace  \left|i^\prime+j^\prime \right| = 2, 
\enspace  {\rm for}  \enspace m = 3.
\end{eqnarray}

\begin{figure*}[th]
    \centering
    \includegraphics[width=0.8\textwidth]{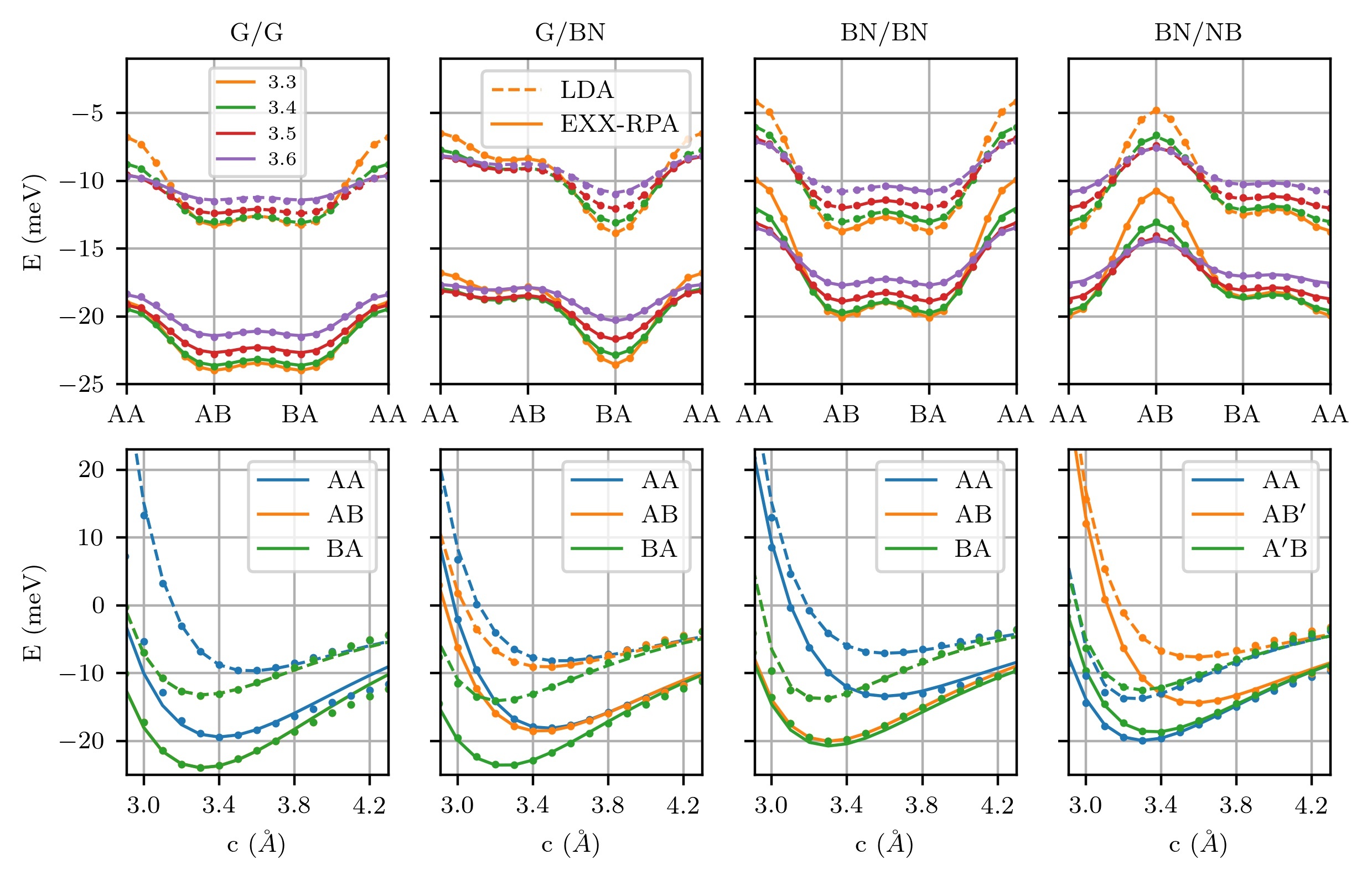}
    \caption{(color online) Comparison of DFT and MD total energies for different interlayer distances and stacking for zero twist angle and lattice-matched $a = 2.48\ \AA$ commensurate geometries of graphene on hexagonal boron nitride as a midpoint between the value of $a = 2.46\ \AA$ used for G/G systems and $2.5\ \AA$ for BN/BN systems. The solid lines fit to the EXX-RPA and the dashed lines reproduce the LDA data. The reference datasets are from the parametrization given in Ref.~\cite{PhysRevB.96.195431}.
    The circles are the MD total energy data points.
    }
    \label{KLIFFFit}
\end{figure*}

%
%
Minimum area commensurate cells for $m=1$ imply that the commensurate supercell period is the same as
the moire period, whose small angle approximation is given by $\ell_{\rm M} \simeq a/ \theta$. 
Because the supercell size grows as the layers are brought to closer alignment
these minimum area supercells will normally be our preferred twist angle choices for our band structure calculations.
Using the current scheme, we can obtain all the smallest commensurate supercells from Ref.~\cite{PhysRevB.81.165105}, as well as additional larger supercells with the same angles.
The expanded set of points can be useful for instance when we 
want to find doubly commensurate real space moire supercells~\cite{leconte2020}. 
In Fig.~\ref{commensuration} We have illustrated with filled and empty symbols the even and odd parity points~\cite{Rode2017}
whose bands are different for large twist angles~\cite{PhysRevB.81.161405}. 
For the typical flat band angles near 1$^{\circ}$ this even/odd signature-differences 
are negligibly small in the limit where the continuum approximation becomes accurate
because both even and odd structures become increasingly similar to each other,
having similar distribution of AA, AB and BA stacking regions.
As a rule of thumb we have even parities when $i, j >0$ or $i, j <0$ and ${\rm modulo}(i-j , 3) = 0$
and we obtain odd parities for the remaining cases.
We note that the parity of the superlattice is reversed for twist angles that 
are equidistant around $30^{\circ}$ since (i) the sixfold symmetry of the triangular sublattice leads to mirror-symmetry around $30^{\circ}$ and since (ii) the $60^{\circ}$-rotation defining the new zero-angle reference inverts the role of even and odd configurations where odd configurations possess coincident lattice positions only for the corner A-A$^{\prime}$ sublattices while the even configurations also possesses an additional coincident site for the B-B$^{\prime}$ sublattice combination.

\begin{table*}[tbh]
\begin{tabular}{|l|l|l|l|l|l|l|l|l|l|l|l|l|}
\hline
\multicolumn{2}{|l|}{DRIP} & $C_0$    & $C_2$    & $C_4$    & $C$       & $\delta$ & $\tilde{\lambda}$ & $A$       & $z_0$   & $B$       & $\eta$    \\ \hline
            EXX-RPA  & CC & 8.568E$-$03 & 1.781E$-$03 & $-$3.277E$-$08 & $-$4.616E$-$02 & 0.465 & 1.259 & $-$4.049E$-$02 & 3.305 & 1.552E$-$02 & $-$1.026     \\ \cline{2-12}
                     & CB & 2.650E$-$02 & 5.326E$-$02 & 7.749E$-$02 & 4.037E$-$08 & 0.881 & 3.055 & 1.544E$-$08 & 3.133 & 8.052E$-$03 & 1.277       \\ \cline{2-12}
                     & CN & 3.585E$-$02 & 1.710E$-$04 & 2.061E$-$02 & 7.224E$-$03 & 0.773 & 3.115 & 5.188E$-$02 & 3.084 & 4.222E$-$03 & 1.083       \\ \cline{2-12}
                     & BB & 0.211 & 9.106E$-$02 & 3.252E$-$02 & 0.232 & 0.939 & 2.834 & 4.292E$-$08 & 2.735 & 3.324E$-$12 & 1.141      \\ \cline{2-12}
                     & BN & 1.446E$-$06 & 0.108 & 0.186 & 1.085E$-$02 & 1.076 & 5.112 & 4.071E$-$08 & 2.871 & 7.748E$-$03 & 2.748    \\ \cline{2-12}
                     & NN & 1.176E$-$02 & 4.701E$-$03 & 8.515E$-$03 & 4.293E$-$02 & 0.779 & 1.310 & 0.197 & 2.958 & 1.787E$-$03 & 2.073     \\ \hline
            LDA  & CC & 5.889E$-$02 & 2.150E$-$02 & 5.265E$-$02 & 1.601E$-$11 & 0.760 & 3.987 & 1.550E$-$02 & 2.988 & 1.181E$-$04 & 1.791    \\ \cline{2-12}
                     & CB & 3.811E$-$02 & 3.606E$-$02 & 0.105 & 2.903E-06 & 0.875 & 5.291 & 3.243E$-$02 & 2.931 & 2.691E$-$03 & 1.147             \\ \cline{2-12}
                     & CN & 6.882E$-$02 & 2.227E$-$02 & 3.967E$-$02 & 1.589E$-$07 & 0.743 & 3.007 & 7.131E$-$07 & 2.941 & 1.093E-02 & 1.268        \\ \cline{2-12}
                     & BB & 0.274 & 0.142 & 3.252E$-$02 & 8.587E$-$07 & 0.696 & 3.121 & 9.474E$-$07 & 2.677 & 3.662E$-$10 & 1.120        \\ \cline{2-12}
                     & BN & 8.727E$-$08 & 0.245 & 0.302 & 7.015E$-$02 & 1.225 & 5.929 & 5.785E$-$08 & 2.834 & 6.142E$-$03 & 2.950      \\ \cline{2-12}
                     & NN & 2.002E$-$02 & 1.658E$-$02 & 1.182E$-$02 & 2.889E$-$03 & 0.774 & 1.323 & 0.159 & 2.535 & 1.111E$-$03 & 1.522    \\ \hline
\end{tabular}
\caption{Fitting parameters for the DRIP functional given in Eq.~\ref{DripFunction} to reproduce either the EXX-RPA or the LDA behavior of G-G, G-BN, BN-BN and BN-NB interlayer interactions. The fitting parameters for B-N interactions accurately fit the  BN-BN and BN-NB (where one of the hBN layers is rotated by $180^{\circ}$) layered materials thus illustrating a certain level of transferability. We use the same values for all pairs for \textit{normal}$_\text{cutoff}=3.7$ (LAMMPS-specific value to find the first three nearest neighbors of an atom in order to calculate the normal to the surface they form), $\rho_\text{cut}=1.562$ and $r_\text{cut}=12.0$. The latter two are potential-specific values that are defined in the appendix after Eqs.~\ref{fcxr} and \ref{dihedralEq}. Atomic charges in the MD relaxation are set to $-0.82275 e$ and $0.82275 e$ for N and B respectively, where $e$ is the elementary charge of the electron.}
\label{DripDataTable}
\end{table*}

\subsection{From \textit{ab initio} calculations to pairwise potentials for molecular dynamics simulations}
\label{abinitiotomd}
%
%

Our multi-scale approach feeds from DFT total energy calculations for various interlayer 
stacking and separation distances to obtain the pairwise potential needed 
in the molecular dynamics (MD) structural optimization that we discuss later in Sec.~\ref{TBSect}.
The reference input data are the total energies of graphene and hexagonal boron nitride homo 
and heterostructures based on the LDA and EXX-RPA calculations~\cite{PhysRevB.96.195431}.
We used the
KLIFF tool~\cite{kliff}
to fit this data to the DRIP potential function~\cite{wen2018} which improves upon the 
Kolmogorov-Crespi (KC)~\cite{Kolmogorov_2005, Naik2019, 1908.10399} potential 
by including a dihedral-angle correction accounting for the local curvature due 
to local corrugations of the layers and allows for an improved 
description of both the total energies and forces especially for capturing the 
interlayer stacking dependent total energies. 
The MD interlayer binding energies are expressed as a sum of pairwise interaction potentials 
$\phi_{ij}$ between sites $i$ and $j$ between layers
\begin{equation}
    E_{\rm inter} = \frac{1}{2}  
    \sum_{i} \sum_{j \notin \text{layer}\, i} \phi_{ij}
    \label{Einter}
\end{equation}
where the 1/2 prefactor accounts for the double counting. 
The DRIP pair-wise potential is given by
\begin{eqnarray}
    \phi_{ij} &=& 
    f_c \left( \frac{r_{ij}}{r_{cut}} \right) 
  \left( e^{-\tilde{\lambda} (r_{ij} - z_0)} 
    [ C + f(\rho_{ij}) +  f^\prime(\rho_{ij},{\alpha_{ij}^{(m)}}) ] \right.  \nonumber \\
    &-& \left.  A(\frac{z_0}{r_{ij}})^6 \right)
\label{DripFunction}
\end{eqnarray}
where 
$f_c( r_{ij} / r_{cut} )$ is a cutoff function~\cite{duin2001} reminded in Eq.~(\ref{fcxr}) of the Appendix
where $r_{ij} = \left| {\bm r}_{ij} \right|$, the cutoff distance is 
$r_{cut} = 12~\AA$, the $\rho_{ij}$ is the transverse projected distance, 
and $\alpha_{ij}^{(m)}$ is a parameter related with the three dihedral angles around a given atom. 
The first term within parentheses $f(\rho_{ij})$ that depends on the transverse distance  
captures the stacking-dependence between layers
and is similar to the KC potential~\cite{Kolmogorov_2005}.
The additional dihedral angle function $f^\prime(\rho_{ij},{\alpha_{ij}^{(m)}})$
accounts for the local curvatures of the graphene ripples. 
The second term is a common attractive $r^{-6}$ London dispersion contribution. 
The interatomic position dependent variables $r_{ij}$, $\rho_{ij}$, $\alpha_{ij}^{m}$,
the optimization parameters $\tilde{\lambda}$, $z_0$, $C$, $A$ and 
those listed in Table~\ref{DripDataTable},
and the functions used in Eq.~(\ref{DripFunction}) are defined both in Ref.~\cite{wen2018} 
and can also be found in the Appendix~\ref{dripappendix}.
For completeness we have also obtained the respective interactions between (C)arbon, (B)oron, and (N)itrogen to describe the G/BN and BN/BN interlayer interaction potentials~\cite{PhysRevB.96.195431}.
All our drip potential parameters will be made available as CBN\_RPA.drip and CBN\_LDA.drip files in the LAMMPS potential directory and these can be used with the input file from the drip example folder in LAMMPS.
We illustrate in Fig.~\ref{KLIFFFit} the RPA and LDA parametrization for different layering combinations of graphene and hBN. The solid lines correspond to the EXX-RPA data and the dashed lines give the LDA-inferred data. The corresponding parameters are included in Table~\ref{DripDataTable}.
Comparison of our parameters for hBN with other existing force fields such as the hBN-ILP potential~\cite{Maaravi2017, Leven2016} will be presented elsewhere. 

\subsection{Molecular Dynamics simulations for the atomic relaxations in twisted bilayer graphene}
\begin{figure}[bt]
    \centering
     \includegraphics[width=1\columnwidth]{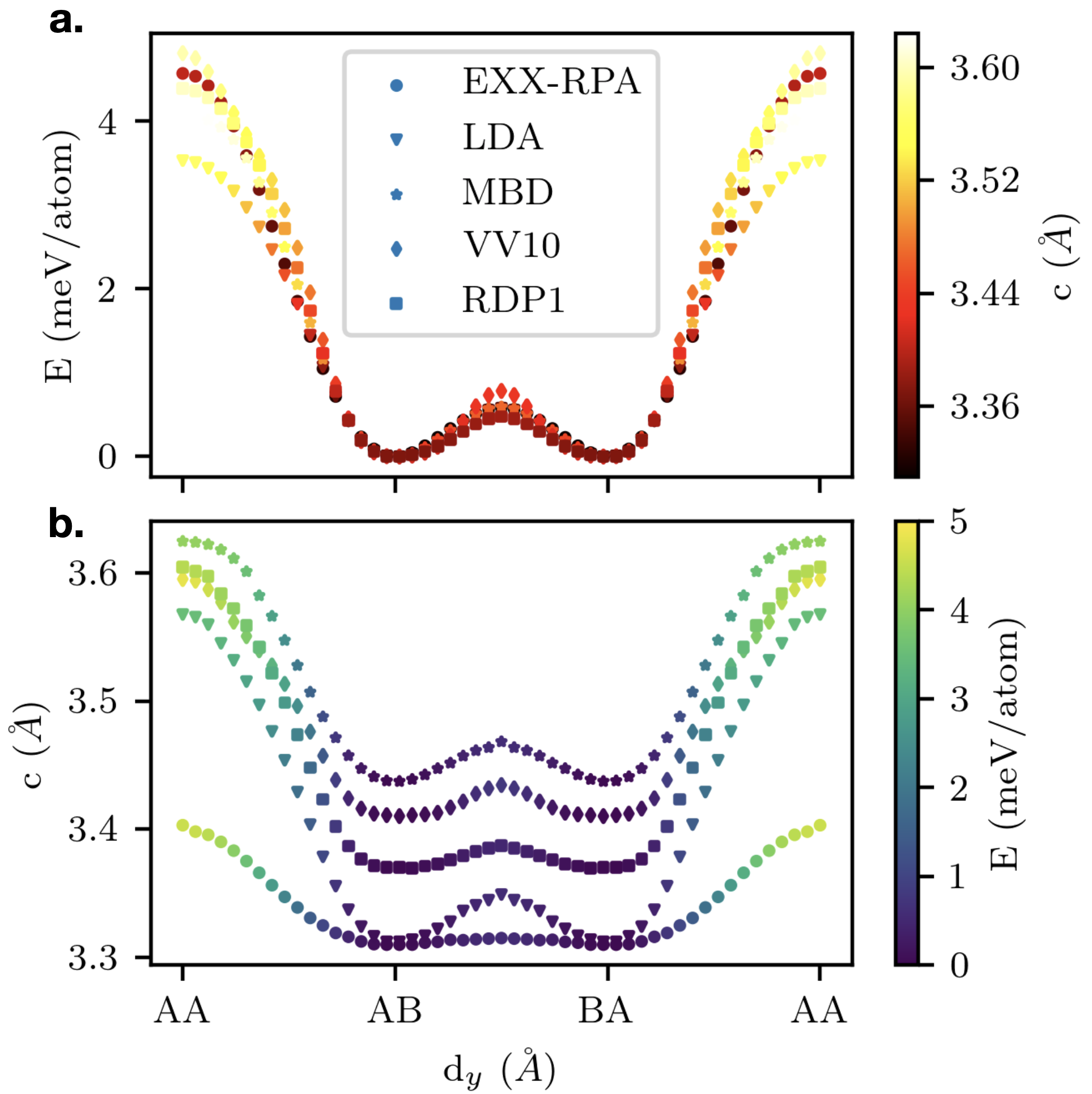}
    \caption{(color online) \textbf{a.} stacking dependent equilibrium energies (AB-stacking set at zero) for the different force fields (see symbols). The colormap illustrates the corresponding equilibrium interlayer distance. We note a mismatch of a factor of about 2 for KC$_\text{VV10}$ with the corresponding curve reported in Ref.~\cite{PhysRevB.87.041108}. \textbf{b.} same information as in the top panel, but the role of the y-axis data and the colormap are reversed. The top panel illustrates that all force-fields predict rather similar interlayer interaction energies, thus explaining the quite similar lattice reconstruction effects. The top panel in contrast shows a large variation in interlayer distances thus explaining the radically different EBSs where the tight-binding models depend strongly on the interlayer distance.
    }
    \label{totalEnergiesFixedInterlayerDistance}
\end{figure}

We perform molecular dynamics simulations using the LAMMPS software 
package~\cite{Plimpton_1995} using different fitted pairwise potentials 
and compare existing implementations against our EXX-RPA and LDA parametrizations for the interlayer interaction energies.
The impact of the MD force field choices in the atomic structure 
is summarized in Fig.~\ref{totalEnergiesFixedInterlayerDistance}
that shows the interlayer distance relaxed stacking dependent total energies
together with the equilibrium interlayer distances as a function of stacking. 
Whereas the stacking dependent total energies remain similar between different
interlayer potential choices we observe that the interlayer distance differences
are reduced in the EXX-RPA approximation by $\sim$0.2~$\AA$ for AA stacking
when compared to the LDA.

For the interlayer stacking dependent total energies
both the KC-type~\cite{Kolmogorov_2005, Gargiulo_2017, van_Wijk_2014}~\footnote{In LAMMPS the \textit{full} implementation without simplifications on the normals must be used to avoid spurious corrugations. We further used the local \textit{https://github.com/sgsaenger/LAMMPS/tree/rdip} branch to avoid a (now corrected) bug existing in the KC \textit{full} implementation existing at the time of writing} and DRIP~\cite{PhysRevB.98.235404} potentials are considered to illustrate their impact on the structural relaxation and changes in the electronic band structure. The RDP1 parameters are taken from the original KC paper with parameters fitted to simulation and experimental data for graphite~\cite{Kolmogorov_2005}, while VV10 uses the same functional form but with parameters fitted to match the VV10 vdW scheme~\cite{PhysRevB.87.041108}. 
The first work on the DRIP potential~\cite{wen2018} was fitted to the many-body dispersion (MBD) scheme~\cite{PhysRevLett.108.236402}, labelled here as MBD. 
As mentioned above, in our work we have fitted the DRIP potential using KLIFF~\cite{kliff} to accurately reproduce EXX-RPA level long-range interactions in bilayer graphene~\cite{PhysRevB.96.195431}, hereafter referred to as RPA. Our LDA data is simply referred to as LDA.
%
%
Intra-layer C-C interactions are described by the REBO2 Brenner potential~\cite{Brenner_2002} and, when hBN is present in the system, intralayer B-N interactions are modeled by EXTEP~\cite{PhysRevB.96.184108}, the extended version of the TERSOFF potential~\cite{PhysRevB.37.6991}. 
We will discuss the impact of the graphene intralayer pairwise potential by comparing against 
the machine learning-informed GAP$_{20}$ potential~\cite{Rowe2020} as well as against the REBO-LB~\cite{PhysRevB.81.205441}
obtained through reparametrization of REBO2.
%
The atomic charges on hBN can be calculated using a Bader analysis based on DFT are $-$0.82275$e$ and 0.82275$e$ for N and B, respectively.
We have neglected the electrostatic interactions between the ionized atoms in our calculations
but these can be included as done for example in h-BN ILP potentials~\cite{Maaravi2017, Leven2016}.
We will show that the choice of the intralayer MD potentials like TERSOFF, EXTEP and REBO2 has a sizeable impact in the MD relaxed geometries and consequently on the associated electronic band-structures.

\subsection{
Intralayer MD potentials and continuum elastic parameters
}
\label{intralayerMD}
%
The relaxed atomic positions result from an interplay  between the interlayer coupling potentials and the intralayer elastic parameters that resist the deformation of the atoms.
%
%
%
Here we report the elastic constants of graphene and hexagonal boron nitride that result from the chose molecular dynamics interatomic potentials that were obtained using the formulas in Ref.~\cite{C8RA02967A} 
for the Young's modulus
\begin{equation}
Y = \frac{C_{11}^2 - C_{12}^2}{C_{11}}
\label{young}
\end{equation}
and the Poisson ratio
\begin{equation}
\nu = \frac{C_{12}}{C_{11}}
\label{poisson}
\end{equation}
where $C_{ij}$ are the elements of the
second order elastic constant matrix 
typically defined as~\cite{C8RA02967A}
\begin{equation}
    C_{ij} = \frac{1}{A_0 c_0}\left( \frac{\partial^2 E_{\rm inter}}{\partial \epsilon_i \partial \epsilon_j}\right)
\end{equation}
where $E_{\rm inter}$ is the energy obtained by summing the pairwise interactions of monolayer graphene (or hBN) as given in Eq.~(\ref{Einter}),
$A_0$ is the equilibrium area of the unit cell of graphene and $c_0$ is the 
out of plane interlayer distance or equivalently the thickness. 
We denote by $\epsilon_i$ the dimensionless strain in direction $i$ that indicates the overall length change ratio. 
In practice, we have used the built-in 
LAMMPS routines to obtain the elastic coefficients by implementing the definitions in Eq.~(\ref{young}~\ref{poisson}),
and these coefficients were obtained by deriving the pressure tensor components calculated by LAMMPS with respect to the strain components~\cite{LammpsManual}. 
We have confirmed the mechanical stability of all these potentials by verifying the following conditions~\cite{Wu2007}
\begin{equation}
C_{44} > 0,\quad C_{11} > |C_{12}|, \quad (C_{11} + 2C_{12}) \, C_{33} >  2 \, C_{13}^2.
\end{equation}
We summarize in Table~\ref{elasticCoeff} the continuum elastic parameters such as the 2D version of the Young Modulus ($Y = Y_{2D}/c_0$), the Poisson ratio and
Lam\'e parameters associated to the C-C and B-N interatomic potentials and how they compare with DFT.
\begin{table}[]
\begin{tabular}{|l|l|l|l|l|l|}
\hline
                     &                  & Y$_{2D}$  (N/m)       & $\nu$        & $\lambda$ (eV/\AA$^2$) & $\mu$ (eV/\AA$^2$)\\ \hline
\multirow{6}{*}{C-C} & DFT (MC) ~\cite{PhysRevLett.102.046808}              & 346 &  0.127   & 3.25    & 9.57  \\ \cline{2-6} 
                     & DFT (GGA)~\cite{Kudin2001}              & 345 & 0.149        & 3.97      & 9.37  \\ \cline{2-6} 
                     & DFT (GGA)~\cite{PhysRevB.80.205407}              & 348 & 0.169        &   4.74    & 9.29   \\ \cline{2-6} 
                     & REBO2~\cite{Brenner_2002}             & 243        &     0.397         & 20.96     & 5.42  \\ \cline{2-6} 
                     & REBO-LB~\cite{PhysRevB.81.205441}         &    364        &    0.098          &      2.51     & 10.34       \\ \cline{2-6} 
                     & GAP$_{20}$~\cite{Rowe2020}    & 317        & 0.193        & 5.22      & 8.31   \\ \hline
\multirow{5}{*}{B-N} & DFT (LDA)~\cite{Sachs:2011jp} & 290       &  0.160          &    3.7       & 7.8       \\ \cline{2-6} 
                     & DFT (GGA)~\cite{Sachs:2011jp} & 284       &  0.153           &    3.4       & 7.7       \\ \cline{2-6} 
                     & DFT (GGA)~\cite{PENG201211}     & 278        & 0.225 &    5.79       &   7.08    \\ \cline{2-6} 
                     & TERSOFF~\cite{PhysRevB.37.6991}          &    250   & 0.31       & 9.81     & 5.96 \\ \cline{2-6} 
                     & EXTEP~\cite{PhysRevB.96.184108}            &  269        & 0.179       & 3.98     & 7.12 \\ \hline
\end{tabular}
\caption{Elastic parameters of different intralayer potentials including the 2D Young's modulus ($Y_{2D}=Y c_0$), the Poisson ratio ($\nu$) and the Lame parameters ($\lambda$ and $\mu$). MC stands for Monte Carlo calculations that allow to go beyond the quasiharmonic approximation. The MD force-field values might differ slightly from published values as we report here the ones calculated directly from a LAMMPS calculation using the script we provide to obtain the different elastic constants for hexagonal systems where the elastic constants are calculated from the coefficients as given in Ref.~\cite{C8RA02967A} using the zig-zag chirality. Experimental values for hBN have been reported in Ref.~\cite{Song2010} with $E \approx 220 - 510$ N/m.}
\label{elasticCoeff}
\end{table}

\begin{figure}[tb]
    \centering
    \includegraphics[width=0.45\textwidth]{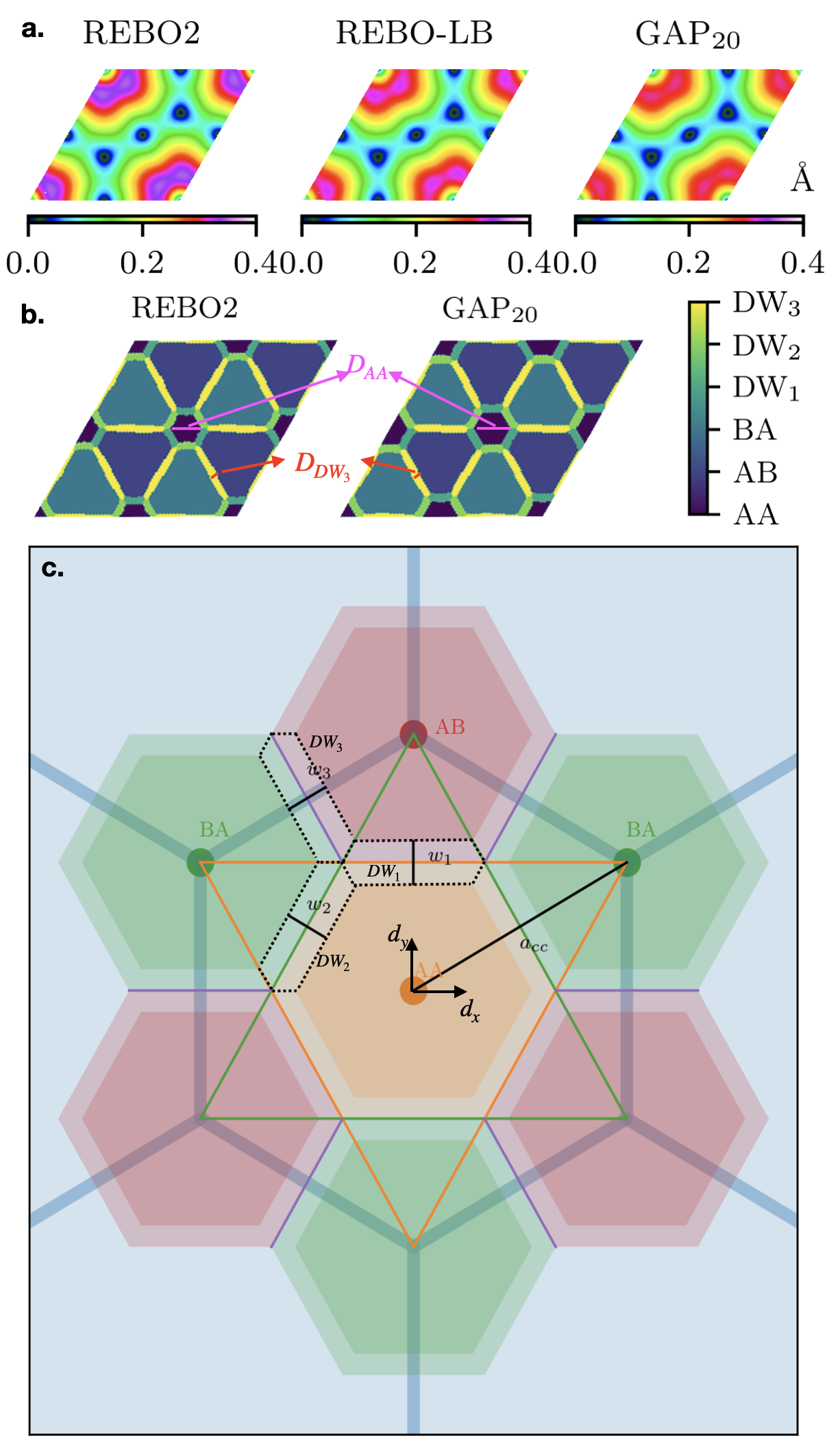}

   \caption{(color online) 
   {\bf a.}  Colormaps of in-plane displacement magnitudes for a tBG with $\theta = 0.53^{\circ}$ for different choices of intralayer potential and same interlayer EXX-RPA potential. The GAP$_{20}$ model gives the largest Young's modulus of the three and is closest to DFT predictions, see Table~\ref{elasticCoeff}, 
   {\bf b.} Comparison of local stacking configuration maps obtained using REBO2 and GAP$_{20}$ for tBG with $\theta = 1.1^{\circ}$ that shows larger strains, with reduced AA stacking regions and expanded AB and BA regions, in the former due to the reduced stiffness. {\bf c.} Graphical representation of the conventions used to classify the local stacking configurations in panel {\bf b.} We distinguish the hexagonal regions around the symmetric stacking AA, AB, BA and the three domain walls regions.
   We have set the three domain wall widths to be $\omega_i = a_{\rm CC}/6$, 
   a choice that leads to a domain wall versus well defined local stacking surface ratio in Eq.~(\ref{sarea})
   $ A_{\rm DW} / A_{\rm supercell} \simeq 30 \%$.
}
    \label{domainWallPlusSketch}
\end{figure}

%

The choice of the interatomic potentials makes a significant impact in the final atomic positions.
Even though the empirical REBO2 and hBN TERSOFF potentials are computationally efficient
they tend to overestimate the strains.
We illustrate this behavior in Fig.~\ref{domainWallPlusSketch}{\bf a.} through the in-plane strain profiles in tBG for 
increasing Young's moduli where
we have compared the lattice reconstructions obtained using the GAP$_{20}$ potential~\cite{Rowe2020}, 
and the REBO-LB potential~\cite{PhysRevB.81.205441}. 
The advantage of GAP$_{20}$ is that it yields a Young's modulus (317 N/m) and a Poisson ratio (0.193) that 
approaches the DFT estimates slightly better than REBO2, while EXTEP matches the experimental G-BN lattice mismatch better than TERSOFF~\cite{PhysRevB.37.6991}. REBO-LB is computationally less expensive than GAP$_{20}$ and reproduces certain DFT elastic predictions~\cite{PhysRevLett.102.046808,Kudin2001, PhysRevB.80.205407} quite well~\cite{Akinwande2017}.
However, this potential becomes unstable for quite small deformations that we could see for example in the resulting elastic coefficients and therefore we have avoided its use.
The in-plane displacements are gradually decreased when switching the C-C interactions from the original REBO2, to the reparametrized REBO-LB and the GAP$_{20}$ ML potential. 
We note that GAP$_{20}$ reduces the maximum displacements from $\sim 0.4$~\AA\ to $\sim 0.3$~\AA\ by about 25\%.

The relative area distribution of different local stacking configurations is also an indicator of the strain effects. 
We break down the moire superlattice area 
\begin{eqnarray}
A_{\rm supercell} &=& A_{\rm stacking} + A_{\rm DW} \label{sarea} \\
&=& A_{\rm AA} + A_{\rm AB} + A_{\rm BA} + 3 (A_{\rm DW1} + A_{\rm DW2} + A_{\rm DW3}) \nonumber
\end{eqnarray}
can be divided into AA, AB, BA local stacking geometries and the areas of the squashed hexagons in the domain wall are multiplied by three to account for the three interfaces. For rigid systems without strains all three local stacking regions and domain wall areas are the same. 
This procedure is reminiscent to building the Wigner-Seitz cells but here we additionally
introduce the domain walls in between the symmetric stacking configurations.
We use the local stacking ${\bm d}({\bm r})$ vector at each unit cell position ${\bm r}$ that we define as the in-plane distance between closest same sublattice atoms in neighboring layers. 
We mark the symmetric AA, AB, BA stacking regions through hexagon tesselation where we choose equal domain wall 
widths of $\omega_i = a_{\rm CC}/6$ as we schematically illustrate in Fig.~\ref{domainWallPlusSketch}c.
With this choice of domain wall width the ratio of relative area of the domain wall versus total supercell area is $A_{\rm DW}/A_{\rm supercell} = $  defined local stacking surface ratio is of 30.5\%.
This definition is used to obtain the evolution of domain wall widths between the AB and BA regions, defined here as $D_{DW_3}$
and we obtain values of $1.3$, $1.6$, $1.4$ and $1.0$~nm for $0.1^{\circ}$, $0.3^{\circ}$, $0.5^{\circ}$ and $1.1^\circ$, respectively, for both GAP$_{20}$ and REBO, suggesting the choice in potential doesn't matter too much for this specific observable.
The trends are similar to the results reported in Ref.~\cite{2008.09761}
that uses different conventions to define the local stacking regions
where the experiments predict a decreasing domain wall widths 
with increasing angle where our range of simulation also allows to locate the an initial increase between $0.1^{\circ}$ and $0.3^{\circ}$ twist angles.  
The diagonals of the hexagonal AA regions are given here by $D_{AA}$ and they are equal to
$3.4$, $4.4$ and $4.5$~nm for GAP$_{20}$ and $3.3$, $3.5$ and $3.9$~nm for REBO2, for $0.3^{\circ}$, $0.5^{\circ}$ and $1.1^\circ$, respectively.
The REBO2 potential gives smaller $D_{AA}$ values, confirming its tendency to overestimate the actual lattice reconstruction with respect to the GAP$_{20}$ potential.

\section{Tight-Binding electronic structure calculations}
\label{TBSect}
Our implementation of the tight-binding electronic structure model for twisted bilayer graphene separates the intralayer and interlayer contributions in an effort to improve its accuracy in a controlled manner. 
A commonly used two center (TC) model~\cite{nam2017, laissardire2012} is based on the interatomic distance vector ${\bm r}_{ij}$ between atoms $i$ and $j$ under the Slater-Koster~\cite{PhysRev.94.1498} approximation, and which 
captures both intralayer and interlayer contributions simultaneously through 
\begin{equation}
    t_{ {\rm TC}, \, ij} = V_{pp\pi}(r_{ij}) \left[1 - \left(\frac{ c_{ij} }{r_{ij}} \right)^2 \right] + V_{pp\sigma}(r_{ij}) \left(\frac{ c_{ij} }{r_{ij}} \right)^2
\label{TCeq}
\end{equation}
where
\begin{equation}
    V_{pp\pi}(r_{ij}) = V_{pp\pi}^0 \exp\left(-\frac{r_{ij}-a_{\rm CC}}{r_0}\right)
    \label{vpppiEq}
\end{equation}
and
\begin{equation}
    V_{pp\sigma}(r_{ij}) = V_{pp\sigma}^0 \exp\left(-\frac{r_{ij}-c_0}{r_0}\right),
\label{vppsigmaEq}
\end{equation}
where $r_{ij} = \left| {\bm r}_{ij} \right|$ is the magnitude of the interatomic distance 
and $c_{ij} = {\bm r}_{ij} \cdot {\bm e}_z$ is the vertical axis projection along the 
$z$-axis normal to the graphene plane. 
For simplicity, here we have defined a fixed normal vector ${\bm e}_z$ along the $z$-axis 
rather than allowing it to tilt with the local curvature following the surface corrugation. 
The parameter $c_0 = 3.35~\AA$ is the interlayer distance, 
$a_{\rm CC} = 1.42~\AA$ is the rigid graphene's interatomic carbon distance, 
$V_{pp\pi}^0 = -2.7$~eV the transfer integral between nearest-neighbor atoms, 
$V_{pp\sigma}^0 = 0.48$~eV the transfer integral between two vertically aligned atoms
that were fitted to generalized gradient approximation (GGA) data for fixed interlayer distances~\cite{laissardire2012}. The decay length of the transfer integral is chosen as $r_0 = 0.184 a$ such that the next-nearest intralayer coupling becomes 0.1 $V_{pp\sigma}^0$. 
The cutoff for this distance-dependent model is finally set to $4.9~\AA$ beyond which 
additional contributions do not affect the observables anymore~\cite{laissardire2012}.
This form is widely used in the literature~\cite{koshino2019, Andelkovic2019} and we will show that is good at reproducing the largest magic angle in tBG around $\sim 1.1^{\circ}$, 
even though the corresponding nearest neighbor 
effective hopping term obtained by adding all intralayer intra-sublattice terms~\cite{jung2013}
gives an estimate of $-2.45$~eV and thus its associated Fermi velocity of graphene is much smaller 
than experiments or even the LDA. 
%

The improved hybrid tight-binding Hamiltonian that we propose treats intralayer and interlayer hopping terms separately as 
%
\begin{equation}
t_{ij} = 
\begin{cases} 
t^{{\rm intra}}_{ ij} 
& \mbox{if }   i \in \mbox{ layer} j \\ 
t^{{\rm inter}}_{ij} & \mbox{if }  i \notin  \mbox{ layer} j.
\end{cases}
\label{hybrid0}
 \end{equation}
Different tight-binding Hamiltonians can be proposed depending on how we define 
the intralayer and interlayer hopping terms. We consider three models. The first one is the scaled two center (STC) model that uses the same
intralayer Hamiltonian $t^{\rm intra}_{ij} = t^{\rm intra}_{{\rm TC}, \, ij}$ as in Eq.~(\ref{TCeq}) but 
uses a scaling factor $S$ for the interlayer hopping terms 
$t^{\rm inter}_{ij} = S \, t^{\rm inter}_{{\rm TC}, \, ij}$,
where the $S$ parameter is fitted to calibrate the magic angle value for each 
choice of force field, or equivalently relaxed atomic structure. 
The $S$ value can also be modified to obtain the band structures for
arbitrary effective twist angles as we will discuss later.
The scaled hybrid (SH) model uses the so called F2G2 model~\cite{jung2013} for the intralayer Hamiltonian $t^{\rm intra}_{ij} = t^{\rm intra}_{{\rm F2G2}, \, ij}$ to improve 
the accuracy when describing a single graphene layer and maintains the same 
interlayer coupling as the STC. 
Our main proposal for this work is the scaled hybrid exponential (SHE)  
model where the interlayer hopping terms are improved to match the interlayer 
tunneling data from {\em ab initio} 
calculations resulting in
\begin{equation}
t_{{\rm SHE}, \, ij} = 
\begin{cases} 
t^{{\rm intra}}_{{\rm F2G2}, \, ij} 
& \mbox{if }   i \in \mbox{ layer} j \\ 
S \, \exp\left[ {(c_{ij} - p)}/{q} \right] \, t^{{\rm inter}}_{{\rm TC}, \, ij} & \mbox{if }  i \notin  \mbox{ layer} j
\end{cases}
\label{hybrid1}
 \end{equation}
where $p = 3.25~\AA$ and $q = 1.34~\AA$. 
This SHE model will be used hereafter to present our results
and we will generally omit its explicit labeling.
In the following we discuss the improvements made for the intralayer and interlayer tight-binding Hamiltonian terms to better match the {\em ab initio} calculations.
%


\subsection{Intra-layer {\em ab initio} tight-binding models}
\label{intralayer}
The intra-layer hopping terms can be improved by using {\em ab initio} calculation fitted tight-binding models as 
presented in Ref.~\cite{Jung:2014hj} where we can systematically control the range of finite hopping terms. 
Here we use the so-called F2G2 model that includes up to two nearest neighbor inter- and intra-sublattice
hopping terms in single layer graphene that enhances the accuracy of the Hamiltonian in the entire Brillouin 
zone while retaining relative simplicity. 
We adopt a Fermi velocity of
$\upsilon_{\rm F} = \sqrt{3} a^2 t_{\rm eff}/ 2 \hbar = 10^{6} {\rm m/s}$ 
that amounts to an enhancement of the 
nearest neighbor hopping term of
$t_{\rm eff} = -3.1$~eV instead of using the LDA value of 
$t_{\rm eff} = -2.6$~eV in a nearest neighbor only model. 
In our F2G2 model this implies using an enhanced physical nearest neighbor inter-sublattice hopping term 
of $t_0 = -3.5$~eV together with other four additional nearest neighbor hopping terms. 
We note that this effective Fermi velocity is closer but somewhat smaller 
than the typical value of $\upsilon_{\rm F} = 1.05$~m/s, or equivalently $t_{\rm eff} = -3.24$~eV,
for graphene on SiO$_2$ substrates whose electron mobilities are comparable to those of tBG.
Additional corrections to the F2G2 intralayer hopping terms in the presence of bond distortions can 
be added through exponentially decaying terms~\cite{Pereira:2009iz} resulting in
\begin{equation}
t^{{\rm intra}}_{{\rm F2G2}, ij} = t_{{\rm F2G2}, \, ij} \cdot \exp\left[ -3.37 \left( \frac{r_{ij} - r_{0,\,ij}}{r_{0,\,ij}} \right) \right]
\label{realStrain}
\end{equation}
%
%
where $t_{{\rm F2G2}, ij}$ are the intralayer interatomic hopping terms of 
the rigid F2G2 model, and 
$r_{0,\,ij}$ are the rigid lattice's in-plane interatomic distances between $i$ and $j$ atoms.

\subsection{Inter-layer scaled tight-binding model}
\label{factorSection}

\begin{figure}[tb]
    \centering
    \includegraphics[width=1.0\columnwidth]{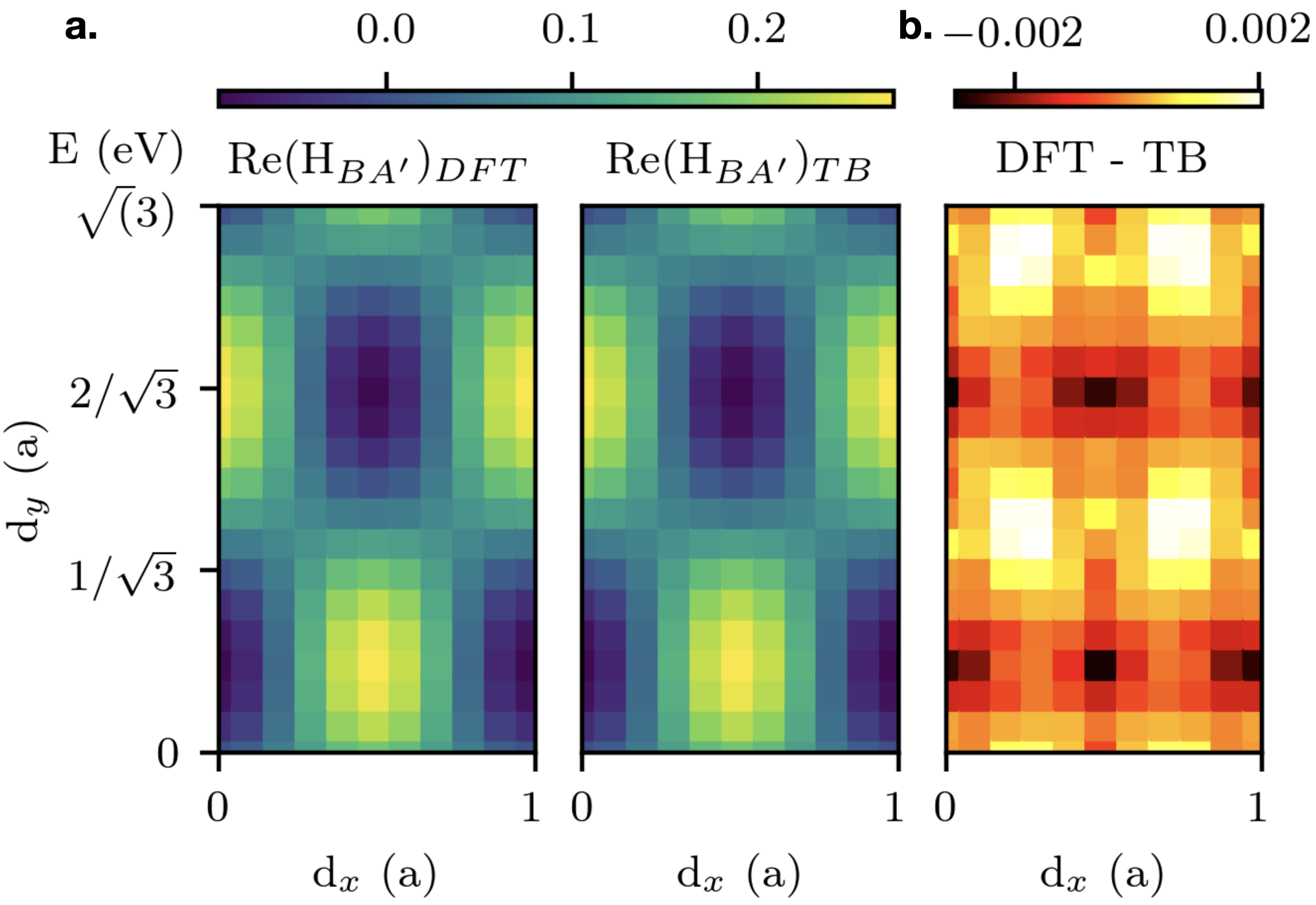}    
    \caption{(color online) \textbf{a.} Comparison of the $H_{BA^\prime}$ tunneling data at $3.5$ \AA\ interlayer distance for different commensurate cell sliding configurations obtained from LDA DFT calculations and the corresponding TB model estimates using the expressions from Eq.~(\ref{STCtunneling}) into Eq.~(\ref{tunnelingEq}). \textbf{b.} Differences between the first two panels are in the meV range and the TB fit can thus be considered accurate for all sliding configurations.}
    \label{ABp} 
\end{figure}
\begin{figure}[tb]
    \centering
     \includegraphics[width=1.0\columnwidth]{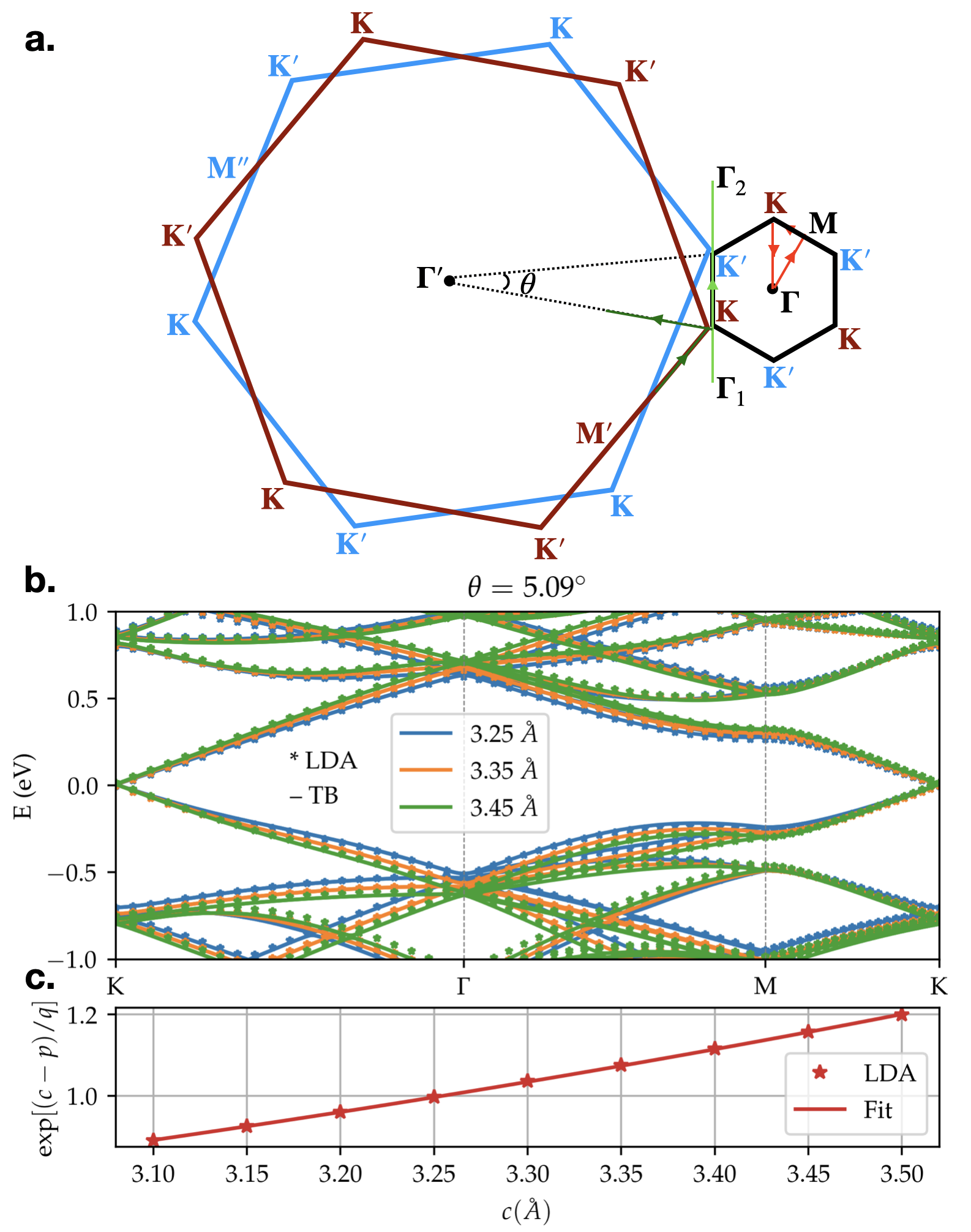}   
    \caption{(color online) 
    (color online) \textbf{a.} Definition of the high-symmetry points used to define the paths of the EBS and the spectral function plots. The EBS are given along the path highlighted in the moire BZ, while the spectral function is centered around the K-point which lies simultaneously in the moire BZ and at the corner of the BZ of the bottom layer of graphene. The branches of the spectral function are calculated towards the $\Gamma^\prime$ and $M^\prime$-points of this same layer of graphene (dark green path) as well as using a straight line through K and K$^\prime$ (light green path) to allow for discussion of bottom and top layer contributions.
    \textbf{b.} LDA DFT vs TB EBS comparisons using the pre-factors given in the inset for different interlayer distances into Eq.~\ref{STCtunneling}. The agreement is satisfying well beyond the $\pm 0.5$~eV range. \textbf{c.} The pre-factors can easily be fit by an exponential dependence, see exponent in Eq.~(\ref{STCtunneling}). }
    \label{AngleBenchmarking} 
\end{figure}

We propose a new interlayer tight-binding Hamiltonian model that allows to better reproduce the ab initio tunneling at the K-point~\cite{jung2014} for different interlayer distances 
and give a better account of surface corrugations or the influence of pressures.
%
The DFT-LDA interlayer tunneling data at the $K$-point was obtained through the 
Quantum Espresso code package using a 
30$\times$30$\times$1 Monckhorst pack k-grid and making minimal modifications to 
Wannier90 code to extract the tunneling~\cite{jung2014}.
The equivalent TB tunneling at the $K$-point between interlayer 
sites of an aligned system
with four atoms per unit cell is calculated using
\begin{equation}
H_{s s^{\prime}}({\bm K}:{\bm d}) = \sum_{j_{s^{\prime}}} t^{\rm inter}_{i_{s} \, j_{s^{\prime}}} 
\exp \left[ {\rm i} {\bm K} 
\cdot ({\bm d} + {\bm r}_{i_{s} \, j_{s^{\prime}}} )
\right]
\label{tunnelingEq}
\end{equation}
where the $s$ and $s^{\prime}$ label the two different sublattices $A$, $B$ and $A^{\prime}$, $B^{\prime}$ of bottom and top layers.
Once we fix a given $i_{s}$ site in the bottom layer sublattice $s$ we carry out a sum over all possible $j_{s^{\prime}}$ sites of sublattice $s^{\prime}$ in the top layer until the sum is converged. Because ${\bm K}$ is a 2D vector, only the in-plane components 
of the 3D distance vector ${\bm r}_{ij}$ contributes in the scalar product. 
The vector ${\bm d} = (d_x,d_y)$ is the sliding of the top layer unit cell dimer with respect to the bottom layer unit cell~\cite{jung2014}. 
We will mainly use the SHE model in Eq.~(\ref{hybrid1}) that contains a local 
interlayer distance $c_{ij}$-dependent exponential function 
\begin{equation}
   t^{\rm inter}_{{\rm model}, \, ij} = S_{\rm model} \, \exp \left[ \frac{c_{ij} -p }{q} \right]  t^{\rm inter}_{{\rm TC}, ij}
\label{STCtunneling}
\end{equation}
%
%
%
%
where $S_{\rm model}$ depends on the relaxation model used.
The parameters in the exponential term were fitted with $p = 3.25~\AA$ and $q=1.34~\AA$ for the rigid twisted bilayer assuming that $S_{\rm model} = 1$. 
This exponential rescaling term alone allows to give an improved agreement 
data where the differences are of the order of 2\% at most. 
See Fig.~\ref{ABp} for the agreement of $H_{AB^\prime}$ maps.
A similar agreement is found for $H_{AA^\prime}$ tunneling although it is not shown.
%
%
%
%
A benchmark comparison of DFT vs our tight-binding implementation for a twist angle 
of $\theta = 5.09^{\circ}$ is shown in Fig.~\ref{AngleBenchmarking} for different interlayer 
distances where we can observe a close agreement between both models especially at low energies.
The DFT-LDA bands were obtained through Quantum Espresso using a
cutoff of 60~Ry (800~eV) with ultrasoft pseudopotentials on a 6$\times$6$\times$1 Monckhorst k-point grid. 
The inset of Fig.~\ref{AngleBenchmarking} shows how this exponential term
enhances the interlayer tunneling with growing $c$ and allowing the TB calculations 
to give bands that agree more closely the LDA-DFT results.
%
%

%

\begin{figure}
    \centering
    \includegraphics[width=0.95\columnwidth]{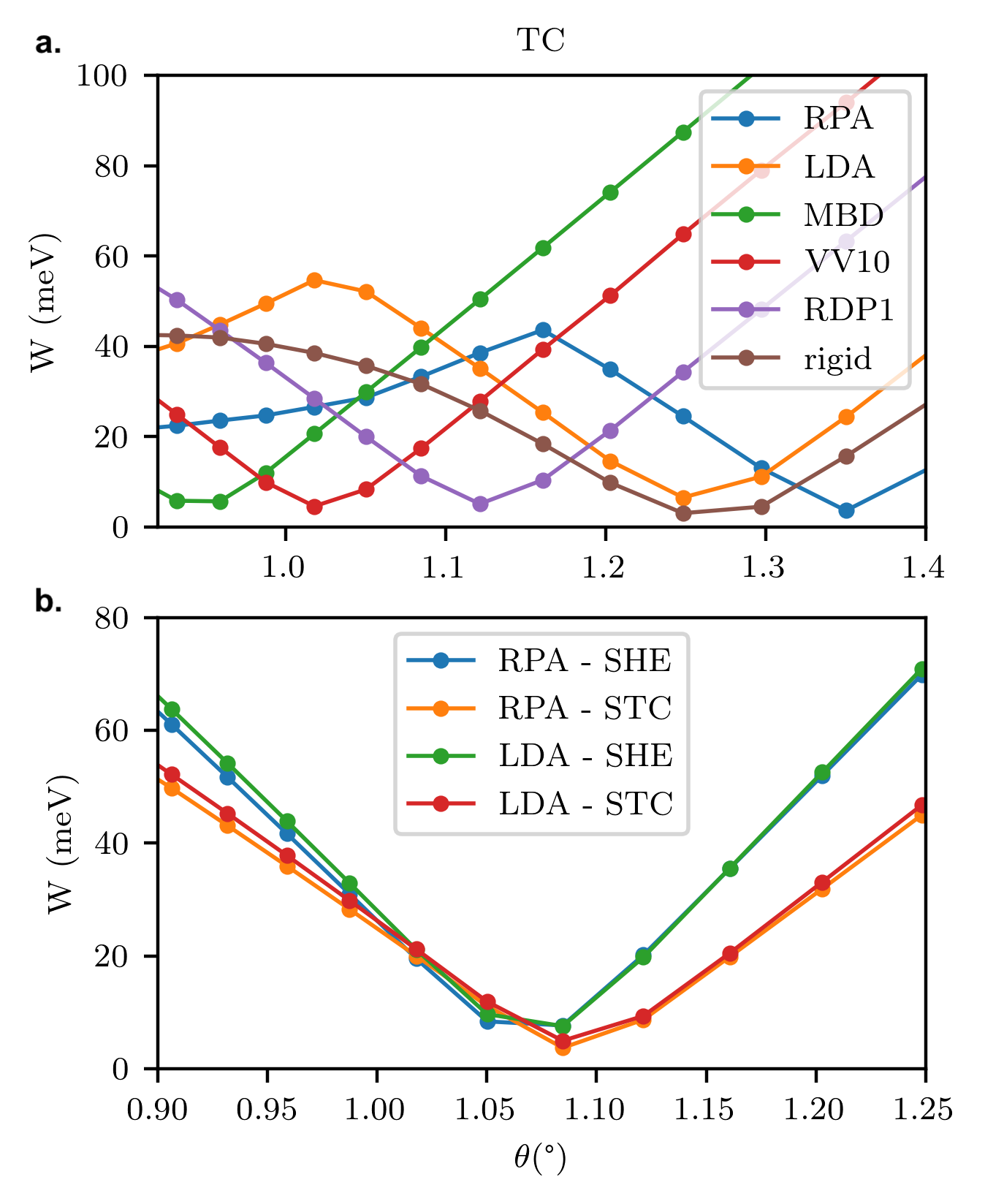}

    \caption{(color online) Bandwidths $W = {\rm max}(E_{\rm cond}) - {\rm min}(E_{\rm val})$ measured as the difference between the maximum and minimum of low energy conduction and valence nearly flat bands for different twist angles and relaxation schemes using REBO2 intralayer potentials. 
    \textbf{a.} Different relaxation schemes either increase or reduce the magic angle values around the rigid system's angle of $\sim 1^{\circ}$. 
    We have used the two center approximation from Eq.~(\ref{TCeq}) for the tight binding Hamiltonian for which the effective nearest neighbor hopping term is equal to $-2.45$~eV. We note that the RPA relaxation increases the magic angle whereas the MBD reduces it. The rigid structure with an interlayer distance of $3.35\ \AA$ leads to a quite large magic when we use this model or around $1.25^\circ$ due to its small Fermi velocity.
    \textbf{b.} Bandwidths with identical magic angle value of $\theta_1 = 1.08^{\circ}$ obtained with our scaled two center approximation (STC) in Eq.~(\ref{hybrid1}) using the calibration parameters of Table~\ref{magicAngleTable}, using an effective nearest neighbor hopping term of $-3.1$~eV. 
    We refer to Fig.~\ref{supplementalBandstructures} in the Appendix for an illustration of the explicit band structures from which some of these bandwidths are extracted.
    }
    \label{Bandwidths_KoshinoSR_3_5}
\end{figure}

The $S_{\rm model}$ prefactor in Eq.~(\ref{STCtunneling}) is a relaxation model dependent parameter that can be varied to calibrate the magic angle to a value of our choice. For brevity of notation we will drop the explicit model label from here onwards.
For convenience we further decompose this fitting constant into two parts as
\begin{equation}
S = \frac{\theta_1 |t_\text{eff}|}{\omega} s = C_1 s
\label{Sfact}
\end{equation}
where we have distinguished the dimensionless $C_1 = \theta_1 \left| t_{\rm eff} \right| / \omega$ 
term and the $s$ is a relaxation model dependent parameter and accounts for the changes in the 
electronic structure due to relaxation strains.
The $C_1$ parameter is the inverse of the dimensionless constant $\alpha_1 = k_{\rm D}/C_1$ corresponding to the first magic angle of tBG
defined in Ref.~\cite{Bistritzer:2011ho}. 
This parameter captures the interdependence between the intralayer Fermi velocity, the interlayer tunneling strength and the magic angle.
The constant $C_1 = 30.4$ corresponding to the first magic angle results from our choice of $\theta_1 = 1.08^{\circ}$, $t_{\rm eff} = - 3.1$~eV and $\omega = 0.11$~eV of rigid bilayer graphene~\cite{Chittari2018} 
and is slightly larger than the value of $C_1 = 27.8$ found numerically in Ref.~\cite{Chittari2018} because 
here we are using different parameters for the Fermi velocity and interlayer tunneling.
The updated scaling parameter $S^{\prime}$ that we obtain when we 
allow for $\delta \theta$, $\delta \upsilon_{\rm F}$, 
$\delta \omega$ parameters to change can be obtained by using the 
updated $C^{\prime}_1$ value
\begin{equation}
S^{\prime} = C^{\prime}_1 s = C_1 \left( 1 + \frac{\delta \theta}{\theta_1} \right) \left( 1 + \frac{\delta \upsilon_{\rm F}}{\upsilon_{\rm F}} \right) \left( \frac{\omega}{\omega + \delta \omega} \right) s.
\label{effectiveS}
\end{equation}
This type of corrections in the scaling parameter $S^{\prime}$ will be convenient for example when we need to calculate the moire band structures of a different effective twist angle 
\begin{equation}
\theta_{\rm eff} = \theta_{\rm ref} + \delta \theta
= S^{\prime} \theta_{\rm ref} / S
\label{effTheta}
\end{equation}
using the reference $\theta_{\rm ref} = \theta_1$ angle of the commensurate superlattice.
Different reference twist angles $\theta_{\rm ref}$ can be used together with a rescaled $S^{\prime}$ parameter to describe an arbitrary $\theta_{\rm eff}$ that varies continuously.
%
%
%
%
%
%
Once $C_1$ is fixed for a given rigid model we can change the strain profile dependent parameter $s$ to calibrate the magic angle of the system.
The $S$ parameters listed in Table~\ref{magicAngleTable} for each type of relaxation strain corresponding to different force fields are chosen to bring the different magic angles to the same $\theta_1 = 1.08^{\circ}$ value as will be illustrated in Fig.~\ref{Bandwidths_KoshinoSR_3_5} of Sect.~\ref{EBSSection}.
The expression for the interlayer tunneling in our SHE model for the EXX-RPA force fields is
\begin{equation}
t^{\rm inter}_{{\rm RPA}, \, ij} = 0.895  
\exp \left[\frac{c_{ij} - 1.34}{3.25} \right] t_{{\rm TC}, \, ij},
\label{ourModel}
\end{equation}
where the coefficient $S = 0.895$ for the RPA relaxed geometries
reduces the overall interlayer tunneling strength with respect to 
the $S \simeq 1$ value of the rigid model
to compensate for the enhancement of the magic angle due to lattice relaxation. 
\begin{figure}[tb]
    \centering
    \label{AAp} 
\end{figure}
In Table~\ref{magicAngleTable} we list the model dependent scaling parameter $S$ needed to bring the 
magic angle to $\theta_1 = 1.08^{\circ}$ for different relaxation models and intralayer 
hopping model choices.



%
\begin{table}[t]
\begin{tabular}{|c|c|c|c|}
\hline
$ S $     &  STC  &  SH  & SHE  \\ \hline
RPA    &  0.752   &   0.951  &   0.895   \\ \hline
LDA    &  0.804   &   1.018  &   0.945   \\ \hline
MBD    &  1.069   &   1.353  &   1.144   \\ \hline
KC-VV10  &  0.970   &   1.247  &   1.091   \\ \hline
KC-RDP1  &  0.884   &   1.136  &   1.018   \\ \hline
Rigid    &  0.856   &   1.083  &   1.008   \\ \hline
\end{tabular}
\caption{
Different $S$ values that we need to multiply in the interlayer 
tunneling term in the TB Hamiltonian 
to bring the magic angle to the experimental value of $\theta_1 = 1.08^\circ$ 
for different relaxed geometries.
The rows are the relaxation models for the atomic structures
and the columns are the tight-binding models for the electronic structures.
The STC model uses the TC intralayer hopping terms of Eq.~(\ref{TCeq}) and 
$S \, t^{\rm inter}_{{\rm TC}, ij}$ for the interlayer tunneling.
The SH model uses the more accurate intralayer F2G2 graphene model
with $\upsilon_{\rm F} =  10^{6}$~m/s corresponding to $t_{\rm eff} = -3.1$~eV and the same interlayer 
tunneling as the TC. 
The SHE model uses the same F2G2 intralayer hopping terms of SH and 
a modified interlayer tunneling with a rescaling parameter $S_{\rm model}$
in Eq.~(\ref{STCtunneling}) that depends on the relaxation model. Similar factors are provided in the Appendix (Table~\ref{magicAngleTableAppendix}) for use when we remove the strain corrections in the F2G2 model of Eq.~(\ref{realStrain}).
}
\label{magicAngleTable}
\end{table}

\section{Electronic structure of lattice relaxed twisted bilayer graphene}
\label{EBSSection}
In this section we obtain the electronic band structure of the tBG models 
from real-space tight binding calculations when the atomic structure is relaxed. 
In particular, we show how the lattice relaxations can impact 
the magic angle value and the shape of the low energy nearly flat bands 
in twisted bilayer graphene. 
The density of states (DOS) as a function of twist angle shows two peaks associated 
each with maxima points stemming either from the valence or conduction bands.
We also show that the a hexagonal boron nitride substrate with a large twist angle with respect to graphene makes a negligibly small impact of the order of a few meV in the band structure.

\subsection{Magic angles of lattice relaxed tBG systems}
Originally the magic angles in tBG continuum models were defined based on the vanishing band dispersion slope at the $K$-point which coincided with the development of almost perfectly flat low energy bands~\cite{Bistritzer:2011ho}. 
This definition becomes less rigorous in a tight binding model
because the bands are not perfectly flat anymore at different 
regions of the moire Brillouin zone and the electron-hole asymmetry becomes more pronounced. 
For this reason we define the magic angles as those twist angles that give rise to the 
narrowest bandwidths of the low energy nearly flat bands
defined as $W = {\rm max}(E_{\rm cond}) - {\rm min}(E_{\rm val})$ 
including both conduction and valence low energy bands.
We will also analyze in the next Section the properties of the low energy bands
from the viewpoint of maxima peaks in the density of states.

In Fig.~\ref{Bandwidths_KoshinoSR_3_5} we show the evolution of the bandwidth $W$ as a function of twist angle for different lattice relaxed atomic structures. 
Within the same two center (TC) approximation given in Eq.~(\ref{TCeq}) 
we notice how the minima positions of $W$ for different relaxation models vary in a wide range of $\theta$ values between 0.9$^{\circ}-$1.3$^{\circ}$ being the rigid model's magic angle around 1$^{\circ}$. 
The rigid model's $W$ does not change much within a twist angle range of $0.04^{\circ}$ as it 
gives rise to a double minima shape, see the appendix Fig.~\ref{Wshape}.
As a general observation, the magic angles are underestimated or overestimated depending on the interlayer and intralayer relaxation schemes while differences in the electronic structure due to intralayer strains as described in Eq.~(\ref{realStrain}) can modify the magic angle typically by about $0.1^\circ$. The pseudomagnetic fields generated by hopping term asymmetries between the nearest neighbors are also dependent on the associated cutoff of the hopping range, i.e. if one only the first nearest neighbor hopping terms are modified the behavior is very similar to the case where one does not include such renormalization at all.

The minima in $W$ as a function of $\theta$ becomes better defined when we account for the lattice relaxations. By using the scaled two center approximation (STC) in Eq.~(\ref{hybrid1})
with the calibration parameters in Table~\ref{magicAngleTable} we notice that all magic angles can be brought to the same 1.08$^{\circ}$ although they have different $W$ values depending on the relaxation model.
The quantitative details in the resulting band structures for the magic angles
are shown in Fig.~\ref{Bandwidths_KoshinoSR_3_5}.
\begin{figure}[t]
    \centering
        \includegraphics[width=0.8\columnwidth]{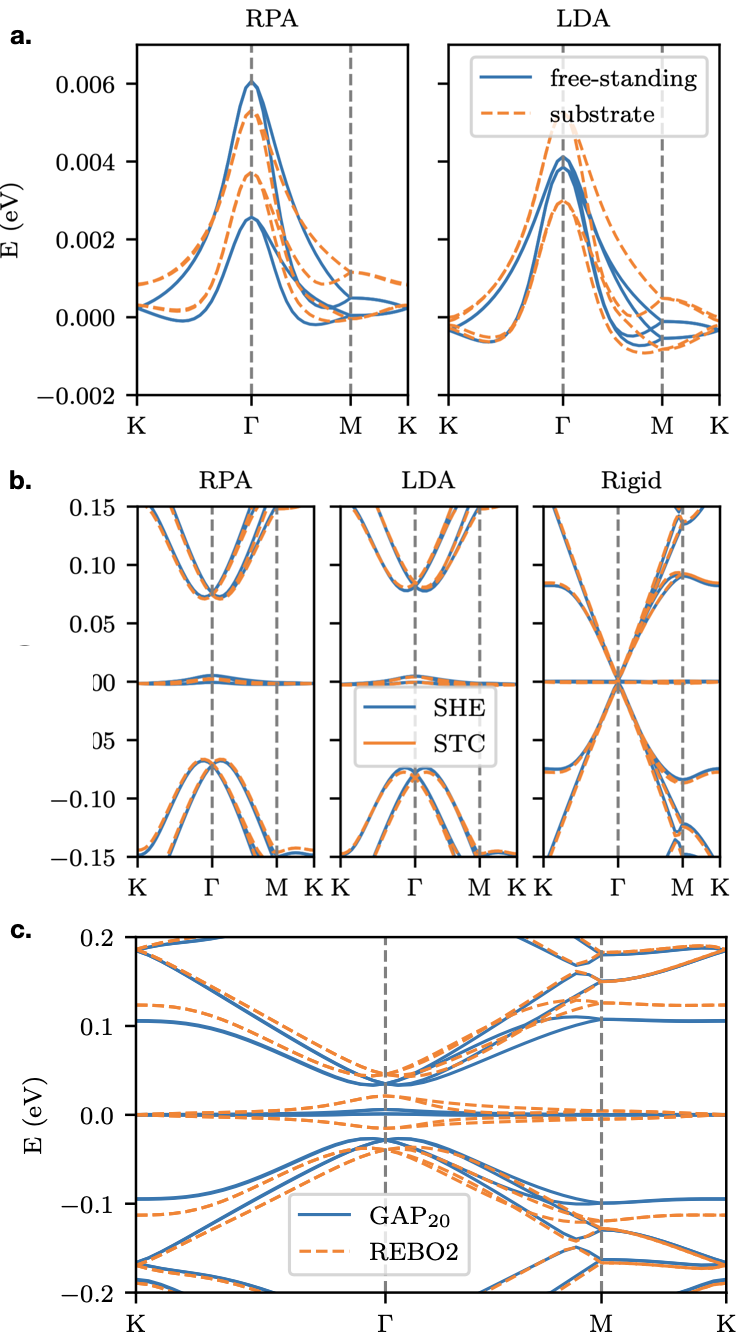}
        \caption{(color online) 
    Electronic structures at $1.08^{\circ}$ for different tight-binding and MD relaxation models
    calculated with the SHE tight-binding model, unless specified otherwise. 
    \textbf{a.} Scaled hybrid exponential (SHE) bands for RPA and LDA relaxation schemes with REBO2 intralayer potentials for free standing tBG in blue solid, and hBN supported and rotated by a large twist angle of 13.37$^{\circ}$.
    We notice that the hBN substrate leads to meV scale modifications in the band structures including a small degeneracy lifting at the K-point of $~1$~meV for the RPA relaxation scheme.
    \textbf{b.} Comparison of the flat bands for different relaxation schemes and tight-binding models. 
    The RPA and LDA relaxation with REBO2 intralayer potentials show similar trends, with the LDA leading to slightly larger band isolation. The impact of the different magic-angle-renormalized TB models are also small with the STC and SHE yielding similar band structures.
    \textbf{c.} Hybrid exponential (HE) bands without prefactor scaling, i.e. the SHE model with $S=1$ in Eq.~(\ref{hybrid1}),
    obtained for REBO2 \textit{vs} GAP$_{20}$ intralayer potentials combined with the EXX-RPA interlayer potentials. The GAP$_{20}$/EXX-RPA relaxation predicts a very flat band when combined with the HE model without further need of prefactor scaling. 
    }
    \label{magicanglebands}
\end{figure}
We further illustrate in Fig.~\ref{magicanglebands} the quantitative impact that different 
relaxation schemes like the RPA and LDA makes in the band structures at the magic angle,
as well as the changes introduced to the atomic structure by an hBN substrate that is
rotated by a large twist angle of 13.37$^{\circ}$ to avoid the strong double moire effects expected 
when G and BN are nearly aligned~\cite{leconte2020}.
We have used up to four hBN substrate layers keeping the bottom layer rigid and verified that 
in practice the result is similar to allowing one hBN layer contacting graphene to relax freely. 
Hence, we can conclude that small differences in the relaxation profile due to the substrate interaction are responsible for the small meV scale differences in the band structures.
In Fig.~\ref{magicanglebands}a. we show that the substrate weakly changes the bandwidth and gaps on the order of $\sim 1$~meV at the $K$-point in the RPA-relaxed system. Because its impact is at the limits of experimental resolution, we argue that the free-standing tBG is a good approximation to tBG deposited on a bulk hBN substrate at a large twist angle. We use this figure to confirm the trend that flat bands tend to be slightly bent at the $\Gamma$-point~\cite{carr2019a}, confirming that already within the non-interacting electron picture 
we can observe a minimum of the valence band away from the $\Gamma$ point~\cite{MacDonaldArpes}.

When we calculated the electronic band structure using the GAP$_{20}$ potential during the relaxation step in Fig.~\ref{magicanglebands}c., we notice that our EXX-RPA interlayer and GAP$_{20}$ intralayer force field combination with the TC model without magic angle calibration already leads to an extremely flat band at $1.08^\circ$. Yet, because REBO2 is still much faster from a computational perspective, this empirical potential is a viable alternative when used in combination with the magic angle scaling factor introduced in Sect.~\ref{factorSection} when our focus is mainly on the electronic structure. For a quantitative analysis of the lattice reconstruction REBO2 falls short according to observations discussed in Sect.~\ref{intralayerMD}.

\begin{figure}
    \centering
    \includegraphics[width=0.95\columnwidth]{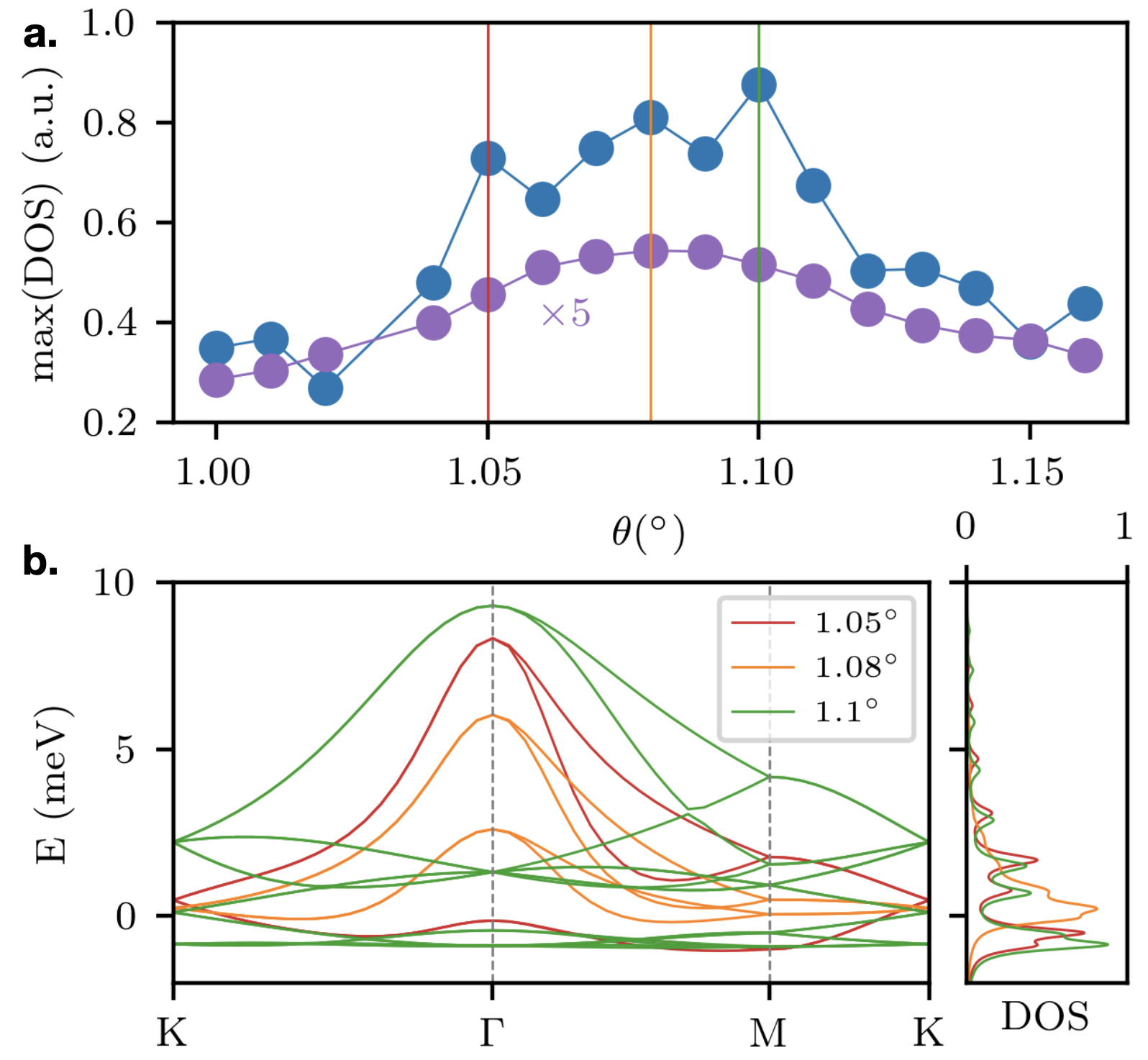}
    \caption{(color online) 
    {\bf a.} DOS vs $\theta$ that shows three peaks at 1.05$^{\circ}$, 1.08$^{\circ}$ and
    $1.10^{\circ}$ when calculated with sufficiently small $\eta = 0.2$~meV in Eq.~(\ref{eqDos}) to resolve the details in the flat bands and a single peak 
    around 1.08$^{\circ}$ when it is increased to $\eta = 3.5$~meV. 
    The magic angle associated to the minimum $W$ corresponds to the single peak value. 
    {\bf b.} The band structures associated to the DOS peak maxima away from the magic angle
    reveal the inherent electron-hole asymmetry and flattening of the valence band over the 
    conduction band. At the magic angle value we find that both electron and hole bands become narrow simultaneously. The associated DOS profiles obtained with $\eta = 0.2$~meV reveal the separation in energy of the valence and conduction bands giving rise to the flat bands away from the magic angle.
    }
    \label{doublepeak}
\end{figure}

\subsection{Multiple DOS peaks structures near the magic angle}
\label{multiplePeak}

In the following we discuss in more detail the band structures of tBG 
calculated near the magic angle and show how the DOS maxima for the valence and conduction bands happen at angles slightly separated from each other when they are calculated with enough energy resolution.
The band structure is represented along high-symmetry points $K-\Gamma-M-K$ 
at selected angles where the supercell BZ is coincident with the moire BZ as discussed earlier. The band structures for twist angles that deviate from this value can be represented in the same moire Brillouin zone by using the updated scaling factor $S^{\prime}$ given in Eq.~(\ref{effTheta}) where an angle difference $\delta \theta$ is added with respect to a reference twist. 
For the electronic band structure calculations, we use exact diagonalization, 
while the density of states are obtained using Lanczos recursion as
\begin{eqnarray}
   {\rm DOS}_{\eta}(E) 
   &=&  - \frac{1}{\pi}\Im m  \left\langle \varphi_{RP} \middle| \frac{1}{E+i\eta - \tilde{\hat{H}}} \middle|\varphi_{RP} \right\rangle  
    \label{eqDos}
\end{eqnarray}
where RP refers to a random phase being used to approximate the trace of large matrices~\cite{lanczos}, $\tilde{\hat{H}}$ is the tridiagonal Hamiltonian approximating 
the full Hamiltonian 
$\hat{H}$ that is useful for continued fraction methods. 
The $\eta$ parameter is the broadening that limits the energy resolution of the DOS.
%
%
The resulting DOS of tBG as a function of twist angle shown in Fig.~\ref{doublepeak}a.
reveals a high DOS region within an interval between 1.05$^{\circ}$ and 1.11$^{\circ}$, 
in other words $\theta = 1.08^{\circ} \pm 0.03^{\circ}$ for a span of twist angles within $\sim 0.06^{\circ}$ around the minium $W$ magic angle.

A closer look at the band structures associated to the twist angles 
in this region shows that the DOS weights are transferred between 
the valence and conduction bands depending on small twist angle changes around 
the magic angle. 
In Fig.~\ref{doublepeak}b. we show that
the DOS weights distribute in both valence and conduction bands at the magic angle 
of 1.08$^{\circ}$ but they are mainly shifted to the valence bands 
when the twist angle is departed by $\pm 0.03^{\circ}$ for 1.05$^{\circ}$ and 1.10$^{\circ}$.

\begin{figure*}[htb]
    \centering
        \includegraphics[width=1.0\textwidth]{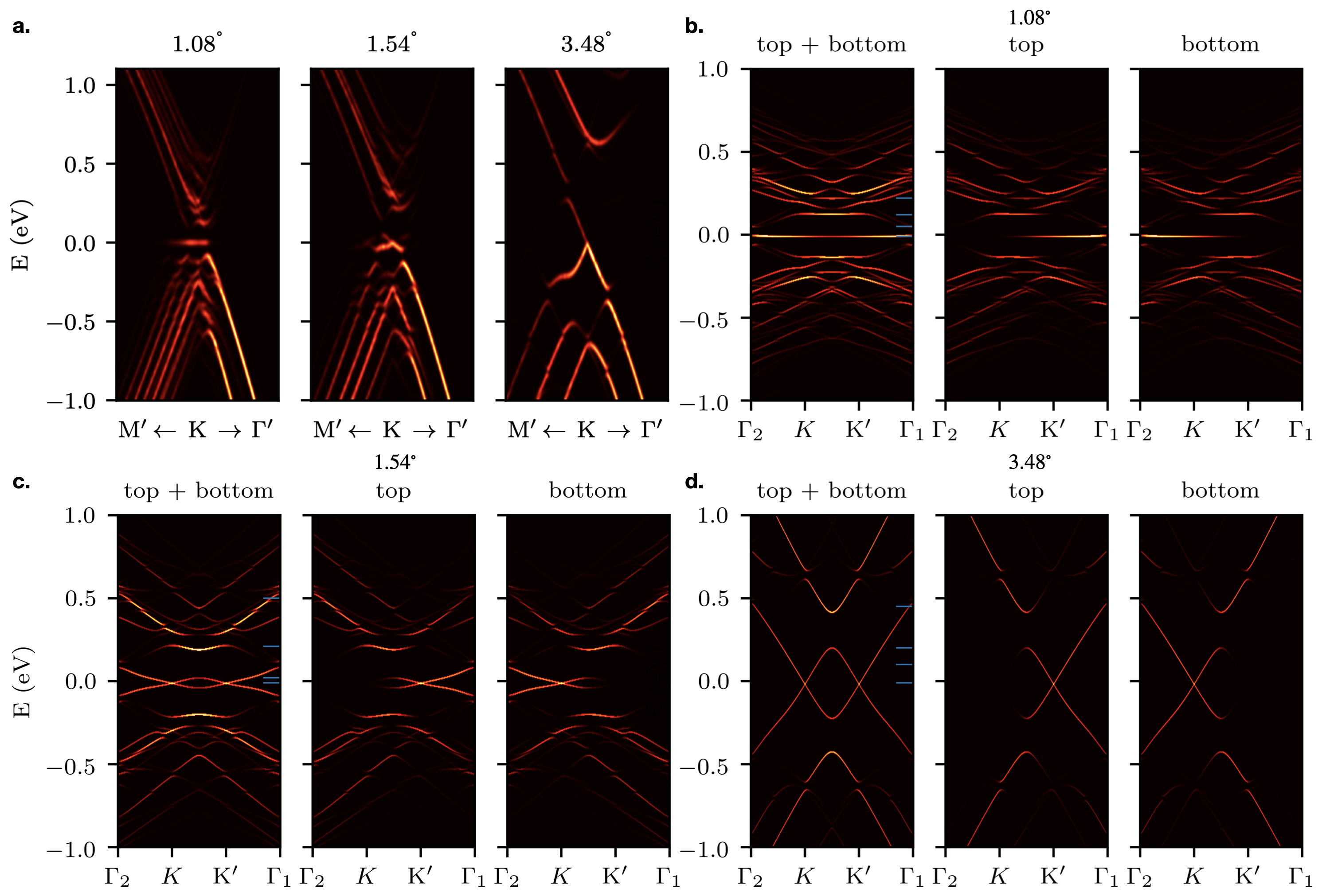}
   \caption{(color online) \textbf{a.} Spectral functions through band-unfolding in the bottom layer for selected angles for the M$^\prime$-K-$\Gamma^\prime$ path in the superlattice BZ where K is also the corner of the bottom graphene BZ. The flat band is clearly resolved in the left panel at $1.08^{\circ}$ twist angle. The middle panel still shows band isolation of the flat bands while the right panel recovers the Dirac cone. The right panel has small secondary Dirac cone signatures around $\pm 0.3$ eV which agrees with the corresponding electronic band structures. 
   \textbf{b.,\ c.,\ d.} Layer-resolved spectral functions through the K and K$^\prime$ points associated to the corners of the bottom and top layers of pristine graphene BZs for $1.08^\circ$, $1.54^\circ$ and $3.48^\circ$ twist angles. The small horizontal lines refer to the energies picked for the 2d-mappings in Fig.~\ref{spectralEnergyCut}. Features from the bottom and top layer are most strongly associated to 
   the K and K$^{\prime}$ points respectively.}
\label{spectralFunctions1}
\end{figure*}

\section{Spectral function calculation}
\label{spectralMethodSect}
The band structure of a superlattice has multiple bands proportional to the atom number in the cell and 
it is often preferable to
obtain the spectral function~\cite{Ku2010, Nishi2017, matsushita2018, PhysRevB.97.165414, MuchaKruczynski:2015ua, MacDonaldArpes, amorim2018} that allows a more intuitive interpretation 
of the quasiparticles as well as a direct comparison against experimental ARPES measurements. 
In our calculations we use the implementation approach in Ref.~\cite{Ku2010} 
to obtain the spectral function of large supercell systems to account for the moire pattern effects. 
Enlarged period supercells are commonly used to account for impurities, vacancies, lattice distortions, or spontaneous long-range orders.
The electronic structure of the zone-folded large supercells can be represented using 
spectral functions in the Brillouin zone of a smaller periodic unit cell through
\begin{equation}
    A_{{\bm k}, n}(E) = \sum_{{ \bm K}J} \left| \langle {\bm k}n \middle| {\bm K} J \rangle \right|^2 A_{{\bm K} J,{ \bm K}J}(E)
\end{equation}
where the $| {\bm K} J \rangle$ eigenbands of the supercell are labeled with capital letters.
The $n$ labels the Bloch function basis $| {\bm k} n \rangle$ with the localized orbital $n$ in the reference small unit cell and can be used to distinguish the layer and sublattice. 
Without loss of generality we chose to represent the $n=1$ orbital content in our spectral functions. 
The $A_{{\bm K} J,{ \bm K}J}(E)$ reduces to a $\delta(E-\epsilon_{{\bm K} J})$ function at the eigenvalue of the superlattice system and  
\begin{equation}
    \langle {\bm k} n | { \bm K} J \rangle = \sqrt{L/l} \sum_{N} w_N e^{i {\bm k} \cdot {\bm R}(N)} \delta_{n,n^{\prime}(N)} \delta_{[\bm k],{\bm K}} \langle {\bm K} N | {\bm K} J \rangle
    \label{spectralEq}
\end{equation}
is a structure factor which is modulated by a position dependent phase term
where ${\bm R}(N)$ is the position of the atom $N$ in the supercell. 
Each term is multiplied by the coefficients of the supercell eigenstate $| {\bm K} J\rangle$ projected in the tight-binding basis $| {\bm K} N \rangle$, and the $L$ and $l$ are the number of k-points in the supercell 
and reference small cell Brillouin zone (BZ) respectively.
The $[{\bm k}]$ denotes the k-point folded into the supercell BZ, and 
here $N$ and $n$ are the orbital indices in the supercell and normal reference cell respectively. 
The weight coefficient $w_N \leq 1$ allows to tune the relative contribution of certain atoms to the spectral function
and allows to improve comparison with experiments since the contribution of the photoelectrons
can depend on which layer is closest to the beam. 
We neglect photon polarization effects that can alter the momentum distribution anisotropy~\cite{PhysRevB.84.125422, PhysRevLett.87.087402}. 
From a computational perspective, the implementation follows closely the scheme of the exact diagonolization band structure calculation. In a first step, one defines the k-point grid $\bm{k}$ in the reference graphene system to be used for the projection. We then use this grid to obtain the corresponding k-points $\bm{K}$ in the moire BZ. We then calculate Eq.~(\ref{spectralEq}) where the parallelization is performed over the $\bm{k}-\bm{K}$ pairs. 
In our implementation, we can rather easily calculate the spectral function of systems containing tens of thousands of atoms within a couple of hours on a node containing 20 cores and 128GB in memory.

In Fig.~\ref{spectralFunctions1}a., we illustrate the resulting spectral functions in Eq.~(\ref{spectralEq}) by projecting in a bottom layer sublattice for a large ($3.48^{\circ}$, right panel) an intermediate ($1.54^{\circ}$, right panel) and the magic angle flat band ($1.08^{\circ}$, left panel) using the SHE model. This type of representation illustrates how the cone from pristine graphene is perturbed by the presence of another graphene layer on top of it. Near the magic angle, we observe the appearance of flat band states that are isolated from the rest of the spectrum
while for larger twist angles, the spectral functions become progressively similar to those of two decoupled graphene layers as suggested by the results in the literature.
The features for $3.48^\circ$ twist angle calculation are qualitatively similar to the features reported for $3.89^\circ$ in Ref.~\cite{Nishi2017}. For the chosen path here centred around the K-point (points on the left and the right are at equivalent distance away from K), we notice a clear asymmetry. Such dark corridor anisotropies in graphene have been linked to the interference between two honeycomb sublattices~\cite{PhysRevB.83.121408, Ulstrup_2015, PhysRevB.93.085409, PhysRevB.77.195403, PhysRevB.51.13614, PUSCHNIG2015193, Bostwick_2007, PhysRevB.84.125422}. Additionally, the presence of interlayer coupling distorts the circular shape that would appear for decoupled graphene layers~\cite{MacDonaldArpes}. Because strain relaxation are present in our simulations, the interlayer tunneling amplitude between orbitals on the same sublattice becomes smaller than the one between orbitals on different sublattices. This in turn has been shown to strongly affect the spectral signatures, even for states that are far away from the flat band~\cite{MacDonaldArpes}.
%

In order to understand the contributions that come from the top and bottom layer, we provide for each of the three angles, a line cut through the $K$ and $K^\prime$ points in panels b., c. and d. as well as constant energy cuts corresponding to the short horizontal blue lines in Fig.~\ref{spectralEnergyCut} in the appendix. The tBG signatures in the left panel is the combination of the bottom and top layer contributions which reside preferentially on the K and K$^\prime$ cones, respectively. Depending on the experimental conditions, one can expect the top layer contributions to appear more clearly in the measurements~\cite{iqbal}.

\section{Conclusion}
\label{conclusion}
We have investigated the interplay between atomic relaxation 
and electronic structure in twisted bilayer graphene (tBG) by combining different molecular dynamics (MD) force-fields and tight binding (TB) models. 
Because different tight-binding and force-field model combinations predict a wide range for simulated first magic angles, we provide a table with coupling strength renormalization pre-factors that match the magic angle at a chosen experimental value of 1.08$^{\circ}$ for a tight binding model with a Fermi velocity of $\upsilon_{\rm F} = 10^{6}$~m/s that is almost 30\% larger than the commonly used two center tight-binding models in the literature.

For the atomic structure modeling part we have provided new parametrizations of a well-established registry-dependent molecular dynamics force fields using the highest level of DFT data available (EXX-RPA) and compared it with existing force fields and associated parameter-sets. We also provide an LDA-parametrized force field for reference purposes. We proposed a way to identify the different local stacking regions through hexagon tessellation of AA, AB, BA stacking regions separated by domain wall regions whose width we set to $\omega_i=a_{cc}/6$. 

The EXX-RPA gives similar energy differences between high symmetry stacking configurations but predicts a smaller overall interlayer difference. This smaller distance leads to a stronger coupling through the distance-dependent tight-binding models, including our improved distance-dependent model parametrization which accurately reproduces LDA commensurate band structures at different interlayer distances. 
We notice that the LDA-inferred force fields in combination with our LDA-fitted tight-binding model also matches quite well the experimental flat bands at around $\sim1.1^\circ$. 
Computationally efficient reactive bond order (REBO) type of force fields tend to underestimate the elastic stiffness of graphene layers overestimating the moire strain profiles.
Albeit being computationally more expensive than most existing readily available empirical potentials, 
the machine-learning potential turns out to provide a good compromise on accuracy and speed when compared to the ab initio molecular dynamics potentials,
while the semi-empirical LCBOPII potential~\cite{PhysRevB.72.214102} that matches well the DFT  
elastic properties might be a good way to simulate very large marginally twisted tBG systems.

For the electronic structure part we have used an intralayer graphene model with up to five nearest neighbor hopping parameters for an improved description of the single layer Hamiltonian using an enhanced Fermi velocity of $\upsilon_{\rm F} = 3\cdot 10^{6}$~m/s. 
The interlayer tunneling was based on a two center approximation to match the {\rm ab initio} DFT calculation data at the Dirac points.
We then used a global prefactor $S$ to modify the interlayer tunneling in what we call the scaled two center (STC)
approximation. Additionally, we refined the tunneling term with an exponential interlayer distance dependent rescaling term
that we named as the scaled hybrid exponential (SHE) approximation aimed at capturing better the layer corrugation effects.
This type of prefactor $S$ calibration is useful for compensating the mismatches of the atomic and electronic structure models.
In fact, the interlayer tunneling of the electronic structure barely required the prefactor calibration
when we used stiffer intralayer force fields in combination with exact exchange and random phase approximation (EXX-RPA) interlayer potentials to obtain a magic angle value close to experiments. 
Adjusting the scaling prefactor is also a useful procedure for obtaining the electronic structure of an effective tBG
twist angle based on a geometry those actual simulation angle is different. It is convenient for example when we need to represent the superlattice bands in the moire Brillouin zone containing a smaller number of atoms.
We have also shown that a hexagonal boron nitride (BN) substrate that makes a large twist angle $\sim 13^{\circ}$ with the contacting graphene layer introduces small lattice distortions in tBG that introduces changes on the order of a few meV in the band structure. 
Depending on the specific choice of lattice relaxation and electronic structure models we observe quantitative differences in the electronic structure near the magic angle like the electron-hole asymmetry of the DOS peaks and the way the low energy bandwidths evolve. 
We noted that peaked density of states associated with the low energy nearly flat bands can be maintained within a twist angle range of 0.05$^{\circ}$ around the calibrated $1.08^{\circ} $magic angle.

In summary, we have taken one step forward towards a systematical improvement in the description of 
the atomic and electronic structure of twisted bilayer graphene.
We expect that a similar approach can be applied to other layered materials by simultaneously controlling the precision of the molecular dynamics force fields and tailoring the tight-binding electronic structure model to reproduce certain well established experimental results or reliable calculation data.

\acknowledgments
We gratefully acknowledge discussions with M. Wen, M. Naik, I. Maity and M. Jain on the molecular dynamics based optimization procedures of the moire strain profiles, and M. I. B. Utama on the analysis of the spectral functions. We thank B. L. Chittari for sharing Born-von Karman relaxation data with us.
N. L. was supported by the Korean National Research Foundation grant NRF-2020R1A2C3009142 and S. J. by grant NRF-2020R1A5A1016518. 
J. A. was supported by the Korean Ministry of Land, Infrastructure and Transport(MOLIT) from the Innovative Talent Education Program for Smart Cities.
J. J. and A. S. were supported by the Samsung Science and Technology Foundation under project SSTF-BAA1802-06. 
We acknowledge computational support from KISTI through grant KSC-2021-CRE-0389, the resources of Urban Big data and AI Institute (UBAI) at the University of Seoul and the network support from KREONET.

\bibliography{all}
\appendix
\renewcommand\thefigure{\thesection.\arabic{figure}}    

\section{Rigid lattice tBG W-shaped bandwidths}
As mentioned in the main text, the rigid geometry exhibits a W-shape, instead of a clear V-shape for the relaxed geometries, when studying the flat-band width with respect to the twist angle. This can be more easily illustrated by plotting this same width, but varying it with respect to the renormalization constant $S$. Within a range $\Delta S$ of about $0.03$, the width is similarly small. We can then use Eq.~(\ref{effTheta}) to estimate that this range corresponds to a $\delta \theta$ range of about $0.04^\circ$, thus showing a relatively large uncertainty on the magic angle value calibration for this rigid system. This W-shape uncertainty in the band width disappears mostly for relaxed systems, however, in those cases the fine features in the DOS calculated with small broadening lead to a wider range of possible magic angles as discussed in Sect.~\ref{multiplePeak}.

 \begin{figure}[h]
     \centering
     \includegraphics[width=1.0\columnwidth]{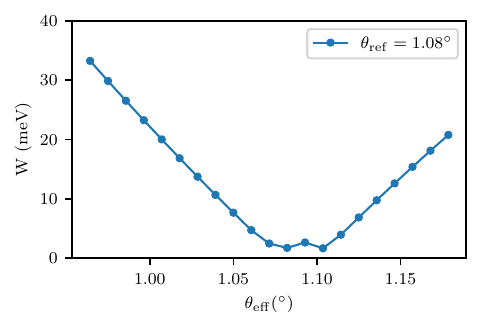}
     \caption{(color online) Evolution of the bandwidth with the magic angle renormalization pre-factor S for the SHE model from Eq.~(\ref{hybrid1}) applied on the rigid system. The minimum for each magic angle curve shows a characteristic W-shape which leads to a small uncertainty of about $0.02$ on the reported pre-factors as the minimization algorithm picks either one of the minima at random. The flat band value is thus unresolved in a range of about $0.04^\circ$.}
     \label{Wshape}
 \end{figure}

\section{DRIP-potential parametrization for G-G, G-BN and BN-BN layered materials}
\label{dripappendix}
\setcounter{figure}{0}

The cutoff function preceding the DRIP pair-wise potential expression, mentioned in the main-text, is defined as
\begin{equation}
f_c(x) = 20 x^7 - 70 x^6 + 84 x^5 - 35 x^4 + 1
\label{fcxr}
\end{equation}
where the cut-off function from Eq.~\ref{fcxr}
with the scaled pair distance $x_r = r_{ij}/r_\text{cut}$ leading to smooth behavior at the $r_\text{cut}=12$ \AA\ cut-off value when used in from of the pair-wise potential in the main-text. 
We remind here also the expressions of the transverse distance function $f(\rho_{ij})$ ~\cite{Kolmogorov_2005} and the dihedral angle function $g(\rho_{ij},{\alpha_{ij}^{(m)}})$~\cite{wen2018} used in Eq.~\ref{DripFunction} for which the new fitting parameters are given in Table~\ref{DripDataTable}:
\begin{equation}
f(\rho_{ij}) = e^{-y^2} \left[C_0 + C_2 y^2 + C_4 y^4 \right]
\end{equation}
with
\begin{equation}
y = \frac{\rho_{ij}}{\delta}
\end{equation}
and
\begin{equation}
\rho_{ij}^2 = r_{ij}^2 - (\bm{n}_i \cdot \bm{r}_{ij})^2
\end{equation}
where the vector connecting atoms $i$ and $j$ is given by $\bm{r}_{ij}$ and the normals to the surface at atom $i$ are calculated at each molecular dynamics step (an early implementation of the KC potential simplified this normal parallel to $z$ but our preliminary tests showed this to lead some unphysical corrugation at zero temperature).
The dihedral function in turn is given by
\begin{equation}
f^\prime(\rho_{ij},{\alpha_{ij}^{(m)}}) = B f_c ({x_{\rho_{ij}}}) \sum_{m=1}^3 e^{-\eta \alpha_{ij}^{(m)}}.
\label{dihedralEq}
\end{equation}
where the cut-off $x_{\rho_{ij}}={\rho_{ij}}/{\rho_\text{cut}}$ is set with $\rho_\text{cut}=1.562$ \AA\ as these 4-body dihedral angle interactions are computationally expensive to calculate. $\alpha_{ij}^{(m)}$ is given by
\begin{equation}
\alpha_{ij}^{(m)} = \cos \Omega_{k_m ijl_1} \cos \Omega_{k_m ijl_2} \cos \Omega_{k_m ijl_3}
\end{equation}
a product of the three cosines of the dihedral angles formed by atom $i$ (in one layer), its \textit{m}$^{th}$ nearest-neighbor $k_m$, atom $j$ (in the other layer), and its three nearest-neighbors $l_1$, $l_2$ and $l_3$.

\section{Electronic band structures}

\begin{figure*}[h]
    \centering
    \includegraphics[width=1.0\textwidth]{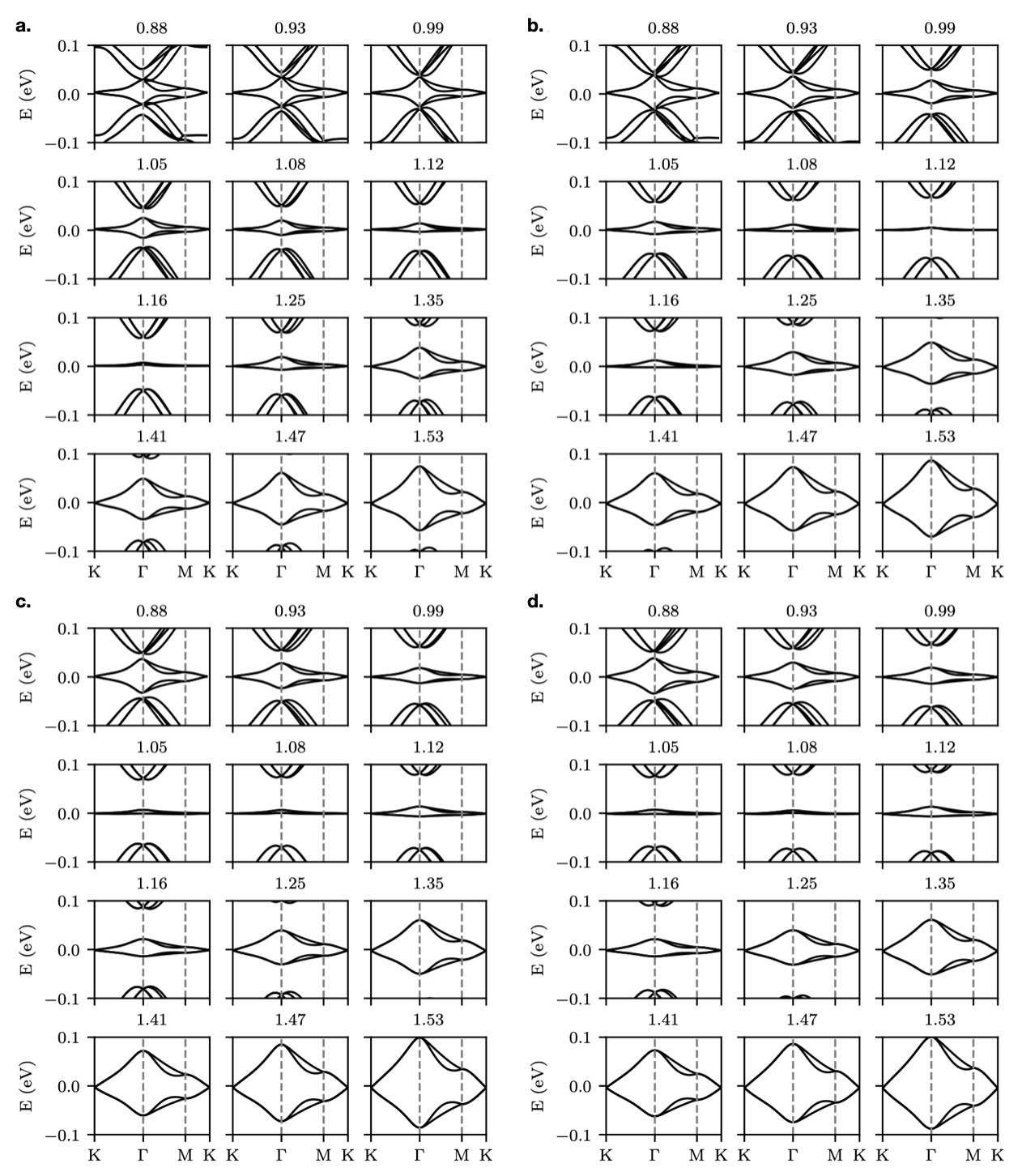}
    \caption{(color online) Band structures for (a) EXX-RPA-informed and (b) LDA-informed force fields using our Hybrid Exponential (HE) TB model without prefactor scaling, i.e. the SHE model with $S=1$ in Eq.~(\ref{hybrid1}), using $t_\text{eff} = -3.1$~eV for the F2G2 model, while \textbf{c.} and \textbf{d.} are the respective counterparts of \textbf{a.} and \textbf{b.} using the SHE model to renormalize the magic angle to $1.08^\circ$.}
    \label{supplementalBandstructures}
\end{figure*}

In Fig.~\ref{supplementalBandstructures}, we compare the band-structures using the Hybrid exponential (HE) TB model without prefactor scaling, i.e. the SHE model with $S=1$ in Eq.~(\ref{hybrid1}) with the band structures using our SHE model where $S\neq1$ with its values given in Table~\ref{magicAngleTable}. Panels \textbf{a.}~and~\textbf{b.} show that the flattest band is different using the RPA or LDA relaxation schemes, for angles at $1.16^{\circ}$ and $1.12^\circ$ respectively.
This shows that the LDA almost does not need to renormalize the $S$ prefactor to match the magic angle to the experimental value. 
When we calibrate $S$ in panels \textbf{c.}~and~\textbf{d.} to give the experimental magic angle $1.08^\circ$ we observe very similar band structures using either of the relaxation schemes. These figures also illustrate how the band-isolation disappears for smaller angles. Well-resolved higher order magic angles will thus be more difficult to define clearly, in agreement with the results reported in the literature~\cite{carr2018}.

\section{Scaling prefactors in a simplified tight-binding model without strain corrections}
\label{scalingFactors}

For reference, we provide the scaling factors from Table~\ref{magicAngleTable} in the main text
for a tight-binding Hamiltonian where we neglect the strain corrections in Eq.~(\ref{realStrain}) for the intralayer F2G2 terms.

\begin{table}[h]
\begin{tabular}{|c|c|c|c|}
\hline
S     & SH   & SHE  \\ \hline
RPA             & 0.867       &   0.809        \\ \hline
LDA             & 0.920       &   0.857        \\ \hline
MBD             & 1.227       &   1.035        \\ \hline
KC-VV10         & 1.107       &   0.972        \\ \hline
KC-RDP1         & 1.018       &   0.917        \\ \hline
Rigid           & 1.083       &   1.008        \\ \hline
\end{tabular}
\caption{Similar scaling prefactors $S$ to calibrate the magic angle to 1.08$^{\circ}$ as in Table~\ref{magicAngleTable} when the intralayer F2G2 model in Eq.~(\ref{realStrain}) does not include the strain effects.}
\label{magicAngleTableAppendix}
\end{table}

\section{Spectral functions energy cuts}

Here we provide additional ARPES simulations to facilitate comparison with experimental energy-cuts of the spectral functions reported in Sect.~\ref{spectralMethodSect}. We chose select energy values for the cuts that are different for each angle as indicated by small blue lines in Fig.~\ref{spectralFunctions1}. The first two rows correspond to the magic angle system at $1.08^\circ$, while the middle two rows and last two rows represent $1.54^\circ$ and $3.48^\circ$, respectively. We separate the plots into three main columns, namely the sum of top+bottom layers, and separate the top and bottom layer contributions. Two small dots indicate the K (lower dot) and K$^\prime$ (upper dot) high-symmetry points as well as the corresponding graphene BZ of the bottom (green) and top (blue) layers. As a general observation, we clearly see that for a same energy, the top+bottom maps are simply the combination of the top and bottom maps. One can fine-tune their respective weight using layer dependent $w_N$ parameters in Eq.~(\ref{spectralEq}) to capture the likely fact that experimental features from the top layer are more easily picked up than the features coming from the bottom layer. Here, we can see that most of the weight of the features coming from the bottom layer are centered around the $K$ point, and the features coming from the top layer are centered around the K$^\prime$ point, as can be expected. For the largest twist angle considered, the K and K$^\prime$ signatures are almost decoupled from each other. We further observe the typical dark corridor anisotropies mentioned in the main text~\cite{PhysRevB.83.121408, Ulstrup_2015, PhysRevB.93.085409, PhysRevB.77.195403, PhysRevB.51.13614, PUSCHNIG2015193, Bostwick_2007, PhysRevB.84.125422}. For higher energies, we also observe concentric features that have been observed in nano-ARPES experiments~\cite{iqbal}.

\begin{figure*}[h]
    \centering
       \includegraphics[width=1.0\textwidth]{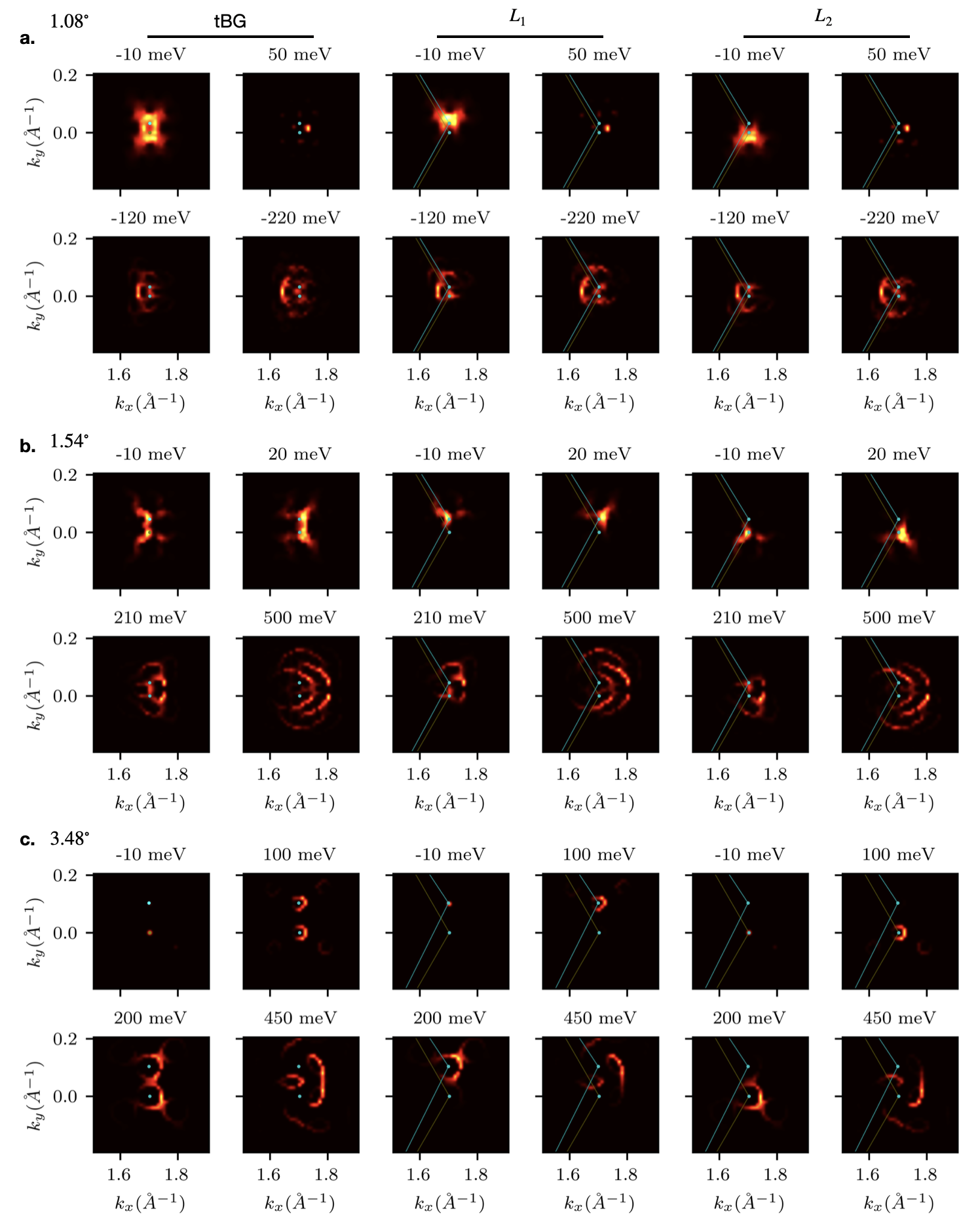}
   \caption{(color online) Spectral function energy cuts for $1.08^\circ$ (\textbf{a.}), $1.54^\circ$ (\textbf{b.}) and $3.48^\circ$ (\textbf{c.}). 
   The first two columns show the total system contributions, the middle two columns are the top layer contributions and the rightmost two columns show the bottom layer contributions for selected energies indicated by the small horizontal lines in Fig.~\ref{spectralFunctions1}.}
    \label{spectralEnergyCut}
\end{figure*}

\end{document}